\documentclass{emulateapj}

\shorttitle{Clustering of Faint UV Galaxies at High Redshift }
\shortauthors{Savoy et al.}

\begin{document}


\title{Keck Deep Fields. IV. Luminosity Dependent Clustering and Galaxy Downsizing in UV-Selected Galaxies at z=4, 3, and 2\altaffilmark{1}}

\author{
Jonathan Savoy\altaffilmark{2}, 
Marcin Sawicki\altaffilmark{2}, 
David Thompson\altaffilmark{3},
and 
Taro Sato\altaffilmark{2}
}

\altaffiltext{1}{Based on data obtained at the W.M.\ Keck Observatory, which is operated as a scientific partnership among the California Institute of Technology, the University of California, and NASA, and was made possible by the generous financial support of the W.M.\ Keck Foundation.}

\altaffiltext{2}{Department of Astronomy and Physics, Saint Mary's University, 923 Robie Street, Halifax, Nova Scotia, B3H 3C3, Canada}

\altaffiltext{2}{Large Binocular Telescope Observatory, University of Arizona, 933 North Cherry Avenue, Tucson, AZ 85721-0065, USA}

\slugcomment{Accepted for publication in ApJ}

\begin{abstract}
We investigate the luminosity dependent clustering of rest-frame UV selected galaxies at $z\sim$ 4, 3, 2.2, and 1.7 in the Keck Deep Fields (KDFs), which are complete to $\mathcal{R}$ = 27 and cover 169 arcmin$^2$.  We find that at $z\sim$ 4 and 3, UV-bright galaxies cluster more strongly than UV-faint ones, but at $z\sim$ 2.2 and 1.7, the UV-bright galaxies are no longer the most strongly clustered.  We derive mass estimates for objects in our sample by comparing our measurements to the predicted clustering of dark matter haloes in the Millennium Simulation.  From these estimates, we infer relationships between halo mass and star formation rate (SFR), and find that the most massive dark matter haloes in our sample host galaxies with high SFRs ($M_{1700}< -20$, or $>50~{M}_\odot$ yr$^{-1}$) at $z\sim$ 3 and 4, moderate SFRs ($-20<M_{1700}< -19$, or $\sim 20~{M}_\odot$ yr$^{-1}$) at $z\sim$ 2.2, and lower SFRs ($-19<M_{1700}< -18$, or $\sim2~{M}_\odot$ yr$^{-1}$)  at $z\sim$ 1.7.   We believe our measurements may provide a new line of evidence for galaxy downsizing by extending that concept from stellar to halo mass.  We also find that the objects with blue UV colors in our sample are much more strongly clustered than those with red UV colors, and we propose that this may be due to the presence of the 2175 \AA\ dust absorption bump in more massive halos, which contain the older stellar populations and dust needed to produce the feature.  The relatively small area covered by the survey means that the absolute values of the correlation lengths and halo masses we derive are heavily dependent on the ``integral constraint'' correction, but the uniformly deep coverage across a large redshift interval allows us to detect several important trends that are independent of this correction. 

\end{abstract}


\section{INTRODUCTION}
The study of high-redshift galaxies is a rapidly developing field that is being driven in part by the ability to efficiently select large samples of galaxies using multi-wavelength photometry.  The Lyman-break technique pioneered by Steidel and collaborators  \citep{Steidel-1999, Steidel-2003, Steidel-2004} working in the rest-frame UV has proven to be very successful at selecting star-forming galaxies at $z>$ 1.5, and at rest-frame optical wavelengths, the $BzK$ selection technique \citep{Daddi-2004} effectively locates galaxies with large stellar populations at $1.4<z<2.5$.  Other high-$z$ galaxy populations include distant red galaxies \citep[DRGs,][]{Franx-2003, vanDokkum-2003}, with $J-K>2.3$, extremely red objects \citep[EROs,][]{Thompson-1999} with large optical to near-infrared colors ($R-K>5$), sub-mm galaxies \citep[SMGs,][]{Smail-2004}, infrared-luminous dust-obscured galaxies \citep[DOGs,][]{Dey-2008} with $F_{24}/F_R>1000$, and strong Lyman-$\alpha$ emitters \citep[LAEs;][]{Hu-1998}.  The large differences in selection criteria ensure that each sample represents a biased view of the total galaxy population at each redshift.  Establishing the properties of these different samples of galaxies, the relationships between them, and their evolution with redshift, is essential to our understanding of how galaxies form and evolve.  

On the theoretical side, it is well established that the clustering strength of dark matter haloes is strongly dependent on mass, with massive haloes being the most strongly clustered \citep{Mo-1996}.  In our current understanding of galaxy evolution, the spatial distribution of different types of galaxies is thought to represent differences in the mass distributions of the dark matter haloes that host them.  For many samples of UV-selected high-$z$ galaxies, an important link has been made between clustering strength and UV magnitude (which is an indicator of star formation), with more luminous galaxies clustering more strongly \citep[e.g.,][]{Adelberger-2005, Lee-2006, Kashikawa-2006}.  This implies a direct link between halo mass and SFR at high-$z$, and it is important to refine this relationship in order to test models of star formation in young galaxies.  Since there is no strong relation between SFR and clustering strength (halo mass) for galaxies at $z\sim0.3$ \citep{Heinis-2007}, it is important to identify the redshift interval over which a relation is in place, and also to determine if it extends to fainter galaxies.  

Clustering studies of high-$z$ galaxies are no longer in their infancy, but many discrepancies and unanswered questions remain.  One of the most significant current limitations in the study of UV-selected galaxy clustering is the lack of uniformly deep UV coverage that probes significantly below the brightest objects at a large range of redshifts.  In particular, most large surveys, such as the Great Observatories Origins Deep Survey \citep[GOODS,][]{Giavalisco-2004}, lack deep $U$ band coverage.  In this respect, the Keck Deep Fields \citep[KDF,][] {Sawicki-2005} provide an important new perspective on galaxy clustering by allowing us to select galaxies with a broad range of UV luminosities ($\mathcal{R}$ = 23 -- 27) in four redshift samples at $z\sim$ 4, 3, 2.2, and 1.7.  The deep UV coverage of the KDFs over a large range of redshifts enables us to search for important trends that have not been previously reported.

As in all of the papers in the KDF series, in the present paper we use the AB flux normalization \citep{Oke-1974} and, unless otherwise noted, adopt $\Omega_M$=0.3,  $\Omega_{\Lambda}$=0.7, $H_0$=70 km s$^{-1}$Mpc$^{-1}$.  To facilitate comparisons to previous studies, correlation lengths are expressed in units of $h^{-1}$ Mpc, where $h=0.7$.

\section{DATA}\label{data.sec}
In this section we briefly summarize some of the most important features of the KDF data on which we base our analysis. A detailed description of the KDF observations, data reductions, and high-$z$ galaxy selection can be found in \cite{Sawicki-2005}, while our $z\sim$4 , 3, and 2 luminosity function measurements are presented and discussed in detail in \cite{Sawicki-2006}.

\subsection{The KDF Survey}
Our analysis is based on results from our very deep multicolor Keck Deep Fields imaging survey carried out using a total of 71 hr of integration on the Keck I telescope. These KDFs use the same $U_n G \mathcal{R} I$ filter set and color-color selection techniques as are used for brighter UV-selected galaxies in the work of \cite{Steidel-1999, Steidel-2003, Steidel-2004}. In contrast to the \cite{Steidel-1999, Steidel-2003, Steidel-2004} work, however, the KDFs reach $\mathcal{R}_{lim}$=27 ; this is 1.5 mag deeper than the Steidel et al. (1999, 2003, 2004) surveys and significantly below  $L^*$ at $z$=2-4: even at  $z\sim$ 4 we reach 2 mag fainter than $M^*$. The KDFs have an area of 169 arcmin$^2$ divided into three spatially independent patches on the sky, with two $\sim$7$\times$10 arcmin$^2$  fields and one $\sim$7$\times$5 arcmin$^2$ field.

To $\mathcal{R}_{lim}$=27, the KDFs contain 427, 1481, 2417, and 2043 $U_nG\mathcal{R}I$-selected star-forming galaxies at $z\sim$4, 3, 2.2, and 1.7, respectively, selected using the \cite{Steidel-1999, Steidel-2003, Steidel-2004} LBG, BX, and BM color-color selection criteria. At our completeness limit, the KDF data probe galaxies with UV luminosities that correspond to SFRs (uncorrected for dust obscuration) of only 1-2 M$_\odot$ yr$^{-1}$. Spectral energy distribution fitting suggests that there is very little dust in galaxies at the faint end of this range \citep{Sawicki-2011}, so these star formation rates are low indeed.  

An important feature of the KDFs is that our use of the $U_n G
\mathcal{R} I$ filter set lets us select high-$z$ galaxies using the
color-color selection criteria defined and spectroscopically tested to ($\mathcal{R}_{lim}=25.5$) by
\cite{Steidel-1999, Steidel-2003, Steidel-2004}.  Our KDF sample is
significantly deeper ($\mathcal{R}_{lim}=27$), making spectroscopic
verification challenging, but the commonality of filters makes us
confident our selection is reliable \citep[for a more detailed discussion, see][]{Sawicki-2005}.  
Nevertheless, we used a combination of published spectroscopy and new deep observations to
verify our selection to the extent possible.  The KDF 09A and 09B fields have been chosen to overlap with the Q0933+2854 field of \cite{Steidel-2003} in which these authors have spectroscopically observed a substantial number of their $U_n G \mathcal{R}$-selected LBGs, and secured redshifts for many of them.  This overlap  makes it possible to verify our selection, at least at the bright end of the KDF sample ($\mathcal{R} < 25.5$). To this end, we cross-matched $19.0 <\mathcal{R} < 25.5$ KDF objects against the \cite{Steidel-2003} catalog of $U_n G \mathcal{R}$-selected LBGs.  The KDF 09A and 09B fields are almost entirely within the Steidel et al.\ Q0933+2854 field, with only 5\% not overlapping (which we excluded from this analysis). In cross-matching, 120 of 139 Steidel objects are recovered in our survey. Several factors are likely responsible for us missing the remaining Steidel et al.\ LBGs: differences in detection masks (e.g., to avoid pixels affected by
bleeding and bright stars); photometric scatter near the $\mathcal{R}
= 25.5$ selection cut; and our crude astrometry, good
to $\sim 1"$ near the edges of survey fields.  Given these
limitations, we essentially recovered all Steidel LBGs in the
overlapping regions. Of the 120 overlapping objects, all of which are classified photometrically as $z \sim 3$ $UG\mathcal{R}$-selected LBGs by \cite{Steidel-2003}, the KDF classifies 54 as $U_n G\mathcal{R}$-selected objects, 44 as BX objects, 3 as BM objects, and none as $G \mathcal{R} I$ LBGs. Furthermore, of these, 33 KDF $U_n G \mathcal{R}$ LBGs have redshifts in the \cite{Steidel-2003} catalog ($<z> = 2.95$) as do 18 KDF BX objects ($<z> = 2.54$).  The differences in classification of objects between \cite{Steidel-2003} and the KDF can be attributed to photometric scatter, particularly in the $U_n$-band: it is reasonable that some BX objects in
our deeper KDF catalog may have been scattered into the $U_n G \mathcal{R}$ selection
in the shallower Steidel et al.\ survey.  The spectroscopic redshift
distributions confirm this interpretation, as the common (KDF-selected) $U_n G \mathcal{R}$ objects
are found near $z \sim 3$ whereas the common (KDF-selected) BX objects have $z \sim
2.5$ --- a slightly higher redshift than the canonical $z \sim 2.2$ of the full BX samples \citep{Steidel-2004} but not surprising given the fact that the \citep{Steidel-2003} spectroscopy was tuned to target $z \sim 3$ galaxies rather than those at lower redshifts.  Furthermore, in addition to matching our objects with the $\mathcal{R}_{lim}<25.5$) spectroscopy of Steidel et al., we have obtained Gemini/GMOS \citep{Hook-2004} spectroscopy for 36 UGR KDF objects, about half of them with $\mathcal{R} >
25.5$.  Continua are detected in all cases, although these are faint.
Their faintness makes it difficult to readily check absorption-line
spectroscopic redshifts, but the spectroscopy has revealed six likely
LyA emitters below $\mathcal{R} = 25.5$ --- i.e., below the Steidel et al. survey limit --- and another six at
$\mathcal{R} < 25.5$.  No evidence of contamination by foreground
interlopers was found.  These spectroscopic tests provide additional
confidence that we are selecting high-redshift objects by extending the
Steidel et al. selection criteria to fainter magnitudes.

As discussed in \cite{Sawicki-2006}, at a rest-frame wavelength of $\sim$1700 \AA, $k$-corrections from suitably-chosen filters are minimal, and the absolute magnitude is very well approximated by
 \begin{equation}
M_{1700} = m_{\lambda_{obs}}  -5 \log (D_L/10 \mathrm{pc}) +2.5\log (1+z).  
\end{equation}
Here, $D_L$ is the luminosity distance and $m_{\lambda_{obs}}$ is the observed magnitude in the $I, \mathcal{R}, G\mathcal{R}$, or $G$ band for $z \sim$ 4, 3, 2.2, or 1.7, corresponding to a rest-frame wavelength of $\sim$1700 \AA.  Our composite $GR$ magnitudes are a simple average of the $G$- and $\mathcal{R}$-band fluxes,
\begin{equation}
G\mathcal{R} = -2.5 \log \left( \frac{10^{-0.4G}+10^{-0.4\mathcal{R}}}{2} \right).
\end{equation}

\subsection{Millennium Simulation}

In order to investigate the halo masses of galaxies sampled by our observations, we make comparisons to the \emph{Millennium Simulation}\footnote{http://www.mpa-garching.mpg.de/millennium/} of dark matter structure formation \citep{Springel-2005}.  The simulation contains 10$^{10}$ particles of mass $8.6 \times 10^8$ $h^{-1} \mathrm{M}_{\odot}$ in a cubic volume 500 $h^{-1}$ Mpc a side, with a resolution of 5 $h^{-1}$ kpc.  The cosmology is a  $\Lambda$CDM model with $\Omega_M=$0.25,  $\Omega_{\Lambda}=$0.75, $h=$0.73, and $\sigma_8=$0.9, similar to that adopted in the rest of this paper.  See \cite{Springel-2005} for a full description.

We investigate the spatial distribution of dark matter overdensities, or ``halos" with $\delta\rho/\rho\sim200$.  Data are available for 64 ``snapshots'' that were saved during the simulation, and from these we select all haloes with masses greater than $\sim$10$^{11} h^{-1}$ M$_\odot$ at $z\sim$ 4, 3, 2.2 and 1.7. 

While the majority of our results utilize the distribution of dark matter haloes in the Millennium Simulation, we also make comparisons to the predictions of the semi-analytical model of galaxy formation described in \cite{DeLucia-2007}, which use the results of the simulation to trace the evolution of dark matter haloes.  Model galaxy catalogs are available for the same 64 ``snapshots'' as the haloes.  For comparison with our observations, we select galaxies based on their star formation rates and total stellar masses.

\section{METHODS}

Galaxy clustering can be estimated using an angular two-point correlation function that measures the number of unique galaxy pairs at a given separation compared to that of a random distribution with the same geometry.  An estimator proposed by \cite{Landy-1993} has become the standard measure of galaxy clustering at high-$z$: 
\begin{equation}
w_{LS}(\theta)=\frac{DD(\theta)-2DR(\theta)+RR(\theta)}{RR(\theta)},
\end{equation}
where 
$DD(\theta)$, $DR(\theta)$,  and $RR(\theta)$ are the number of galaxy-galaxy, galaxy-random, and random-random pairs with angular separations between $\theta-\Delta\theta/2$ and $\theta+\Delta\theta/2$.  Random catalogs were compiled by inserting large numbers of artificial objects with various magnitudes and random positions into the KDF images and then attempting to recover them using the same selection procedure that is used to detect the galaxies.  This ensures that the random catalogues have the same geometry as the galaxy samples, avoiding areas near bright stars and the edges of the images.  To reduce the noise in the random distributions, the $w_{LS}$ calculations are carried out for several thousand different distributions drawn from the random catalogs.

On large scales, the angular correlation function can be approximated by a power law of the form
\begin{equation}
w(\theta)=A_w\theta^{-\beta}.
\end{equation}
Because the total number of random pairs is equal to the number of galaxy pairs, and $w_{LS}$ measures the excess probability of finding a galaxy pair relative to finding a random pair,  the observed correlation function cannot be positive for all $\theta$.  The measured correlation function is reduced by an  amount known as the integral constraint (IC), which increases with clustering strength and decreases with field size,

\begin{equation}
IC \sim \frac{1}{N_{gal}}+\sigma_{w}^2,
\end{equation}
where the first term accounts for Poisson variance and the second term corrects for the fact that the mean galaxy density is estimated from the sample itself and no fluctuations larger than the field size have been taken into account \citep{Peebles-1980}.  The second term can be estimated numerically using the random-random pair counts for the field \citep{Infante-1994, Roche-1999},
\begin{equation}
\sigma_{w}^2=\frac{\sum_i A_w\theta_{i}^{-\beta}RR(\theta_i)}{\sum_iRR(\theta_i)}.
\end{equation}

The amplitude and slope of the angular correlation function are then determined by the fitting function 
\begin{equation}
w(\theta)_{obs}=A_w\theta^{-\beta}-IC.
\end{equation}

The Poissonian errors \citep{Landy-1993} are estimated for the angular correlation function, assuming the weak correlation limit,
\begin{equation}
\delta w _{obs}(\theta)=\sqrt{\frac{1+w(\theta)}{DD(\theta)}}.
\end{equation}
This estimate was confirmed to follow bootstrap errors at separations larger than 10$''$, where the fitting is performed.

To compare with previous studies, and to estimate the mass of the hosting dark matter haloes, it is useful to calculate the spatial correlation function, which is also a power law on large scales, 
\begin{equation}
\xi(r)=\left( \frac{r}{r_{0}} \right)^{-\gamma},
\end{equation}
where $r_0$ is the spatial correlation length and $\gamma = \beta+1$.  The spatial correlation function is related to the angular correlation function through the Limber inversion \citep{Totsuji-1969,Magliocchetti-1999},
\begin{equation}
A_w=\frac{H_{\gamma}r_{0}^{\gamma}\int F(z)r_{c}^{1-\gamma}(z)N^2(z)E(z)dz} {(c/H_0)\bigl[\int N(z)dz\bigr]^2},
\end{equation}
where $r_c(z)$ is the comoving radial distance, $N(z)$ is the survey selection function, and
\begin{equation}
H_\gamma=\Gamma \left( \frac{1}{2} \right) \frac{\Gamma \left[ (\gamma-1)/2\right]}{\Gamma(\gamma/2)},
\end{equation}
 \begin{equation}
E(z)=\sqrt{\Omega_m(1+z)^3+\Omega_\Lambda}.
\end{equation}
$F(z)$ is the evolution of clustering with redshift, which is assumed to be negligible within the samples considered here, and is set equal to 1. 

\begin{figure}
\plotone{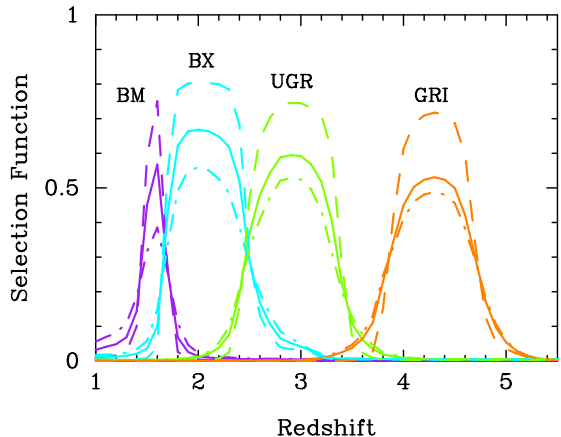}
%\plotone{./figs/sel.ps}
\caption{\label{sel.fig} 
Number-weighted selection functions for the BX, UGR, and GRI samples.  The solid, dashed, and dash-dot lines correspond to magnitude ranges of $\mathcal{R}$ = 23-27, 23-25.5, and 25.5-27, respectively.  %The histogram represents an arbitrary scaling of the spectroscopic results of \cite{Steidel-2004} for galaxies in the BM sample with $\mathcal{R}\le25$.  
\vspace{10pt}}
\end{figure}

The KDF survey selection functions for different redshifts and magnitude ranges are determined using the completeness estimates described in Sawicki \& Thompson (2006).  These simulations use model objects, with colors matching those expected among high-$z$ galaxies, that are added to the data images; the same photometry procedure is then used on these images as on the data-only images to gauge the completeness as well as photometric scatter in the data.  Except at $z\sim 1.7$ our simulations produce redshift distributions that match very well with the \cite{Steidel-1999, Steidel-2003, Steidel-2004} spectroscopic observations at $z\sim$ 4, 3,  and 2.2.  At $z\sim$ 1.7, the modeled selection functions do not accurately reproduce the observed distribution, so we have less confidence in the absolute values of the correlation lengths we report for this sample.  Figure~\ref{sel.fig} shows the number-weighted selection function estimates for various magnitude ranges for the four redshift samples.

\section{CLUSTERING RESULTS}

\begin{figure*}
\plotone{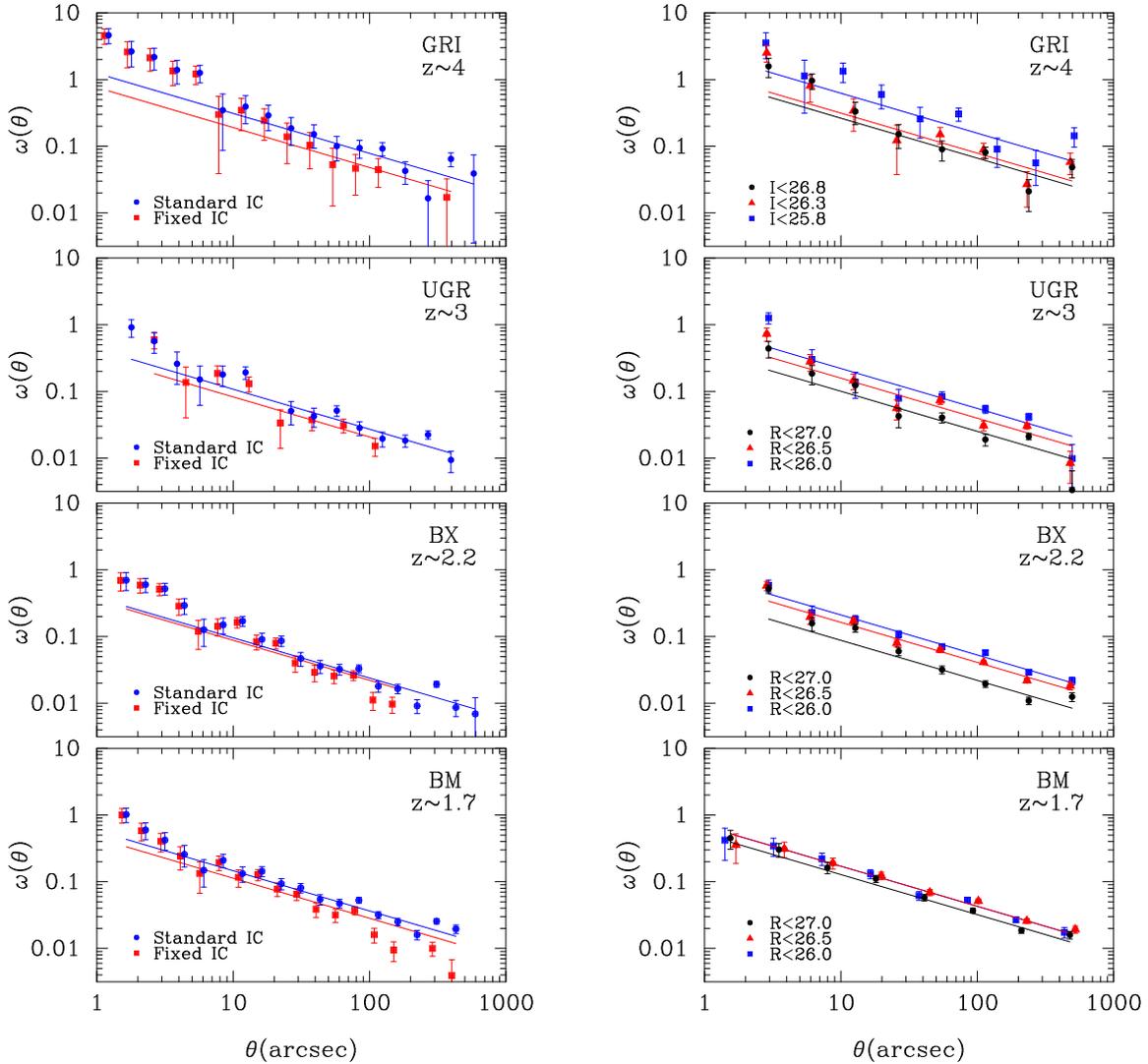}
%\plotone{./figs/wlsall.ps}
\caption{\label{wlsall.fig} 
$a.)$ Angular correlation functions for the complete samples.  Solid lines represent the best fitting power laws to the data between 20 and 700$''$ for the standard IC fitting procedure and from 20 to $\sim$ 200$''$ for the fixed IC fitting procedure.  The data points are offset horizontally for clarity.  $b.)$ Angular correlation functions for various UV magnitude thresholds using the standard IC fitting technique.}
%\vspace{10pt}}
\end{figure*}

It is clear that the integral constraint has a significant contribution to the measured clustering strength of galaxies in relatively small fields.  To assess the importance of the integral constraint, the correlation function is estimated in two ways:  first using the integral constraint calculation as outlined in the previous section (which we refer to as the \emph{standard} IC fit), and another using an assumed value of IC=0.01 (referred to as the \emph{fixed} IC fit), which is consistent with values derived from larger fields \citep{Adelberger-2005, Lee-2006}.  Standard IC values for each sample can be found in Table~\ref{CF.tab}. While fixing the IC produces a somewhat artificial representation of the data, it can give an indication of the impact of this correction on the results and provide a valuable confirmation of general trends that may be suspect in light of possible systematics introduced by the IC.  With this in mind, our analysis will proceed along two parallel paths, using both values of the IC correction.

Since the sample sizes used in this study are generally too small for a robust determination of the correlation function slope, $\beta$,  we assume a value of $\beta$=0.6, which is consistent with studies of LBGs \citep{Adelberger-2005, Lee-2006} and our uncertain estimates.  Additional values of $\beta=$ 0.8 or 0.4 are also explored for some subsamples in order to to facilitate comparisons to the Millennium Simulation. 

The left-hand panels of Figure~\ref{wlsall.fig} show the integral constraint corrected angular correlation functions for both the standard and fixed IC fitting techniques for the four complete samples, along with the best fitting power laws.  For the standard IC fitting procedure, the data is fit between 20 and 700$''$.  The lower limit minimizes the potential contribution of multiple galaxies sharing the same halo, and the upper limit roughly corresponds to the extent of the images.  For the fixed IC fitting procedure, the lower limit is again set at 20$''$, but the upper limit is reduced to $\sim100''$ in order to maintain a power law relation throughout the fitting window.  Data points at scales larger than the outer limit are also plotted for completeness.  Artificially reducing the IC significantly reduces the inferred clustering strength.  The effect is most noticeable in samples with large angular clustering amplitudes (and therefore large IC values), as in the case of the $z\sim$ 4 sample in Figure~\ref{wlsall.fig}.  While the IC does not have a large effect on the complete $z\sim$ 2.2 and $z\sim$ 1.7 samples presented in Figure~\ref{wlsall.fig}, it has a more significant impact on subsamples that are more strongly clustered.

All samples show evidence of an excess on small scales with respect to the large scale power law, which is generally attributed to multiple galaxies residing in the same halo \citep[e.g.,][]{Lee-2006}.  The excess is similar to what has been observed at $z \sim$ 4 by \cite{Lee-2006} and at $z \sim$ 3 by \cite{Hildebrandt-2007}. Ours is the first time a small scale excess has been reported for UV-selected galaxies at $z \sim$ 2.2 or 1.7.

\subsection{Systematic Uncertainties}

Possibly the largest source of systematic uncertainty in the measured correlation functions is the IC correction.  The relatively small field of view of the KDFs requires that large IC corrections need to be applied for the standard IC fitting technique, and the resulting correlation lengths are sometimes significantly greater than those derived from larger fields.  However, if a smaller IC, appropriate to a larger field is artificially implemented, the KDF results are in agreement with those of large area surveys.  This means that even when the uncorrected measurements of two surveys are in agreement, the larger IC corrections for our smaller field can increase the clustering strength we infer by $\sim 20-30 \%$.  This gives an indication of the potential magnitude of the systematics associated with the IC.  It is important to note that while the IC can have a substantial impact on the absolute values of the correlation lengths we derive, the important general trends we observe are independent of this correction.

Also of potential importance are the redshift selection functions, which are needed to perform the Limber inversion. In \cite{Sawicki-2006}, the estimated selection functions at $z\sim$ 4, 3, 2.2 were shown to be in agreement with the Steidel et al. spectroscopic samples,  but the exact distributions, especially for the faint objects,  remain untested.  At $z\sim$ 1.7, the simulated selection functions do not accurately reproduce the Steidel et al. spectroscopic results. The inferred correlation lengths based on the synthetic selection functions are $\sim 30\%$ and $\sim 20\%$ lower for the $z\sim$ 1.7 and $z\sim$ 2.2 samples and $\sim 25\%$ and $\sim 20\%$ higher for the $z\sim$ 3 and $z\sim$ 4 samples, in comparison to the correlation lengths  based on the Steidel et al. spectroscopic results. It is also worth noting that the simulated selection functions show variations with magnitude, which cautions against using the selection functions of the brightest, spectroscopically confirmed members, across all magnitude ranges. These variations generally affect the inferred correlations lengths by less than $10\%$.

Another potential source of uncertainty is due to "cosmic variance" --- i.e., field-to-field variations caused by the fact that on small scales the universe is not homogenous.  The KDF survey consists of three spatially independent sightlines which at least partially guards against the cosmic variance effect.  Indeed, in our luminosity function work (Sawicki \& Thompson 2006a) we found no evidence for significant field-to-field variation among the KDF fields.  Therefore, while cosmic variance remains a potential concern, we do not see evidence for it in our data.  

Contamination of our high-redshift samples by other populations --- most notably lower-$z$ interlopers --- could also affect the clustering results. In this respect it should be noted that spectroscopic observations at intermediate magnitudes, $R$=24--24.5 (Steidel et al.\ 2003, 2004),  find only low interloper fractions of $<$5\% in the $U_nG{\mathcal R}$-selected samples ($z \sim $1.7, 2.2., 3). The interloper fraction is likely to be lower still at fainter magnitudes since the ratio of galaxies to Galactic stars, which are one of the main classes of interlopers, increases towards fainter magnitudes.  Adelberger et al.\ (2005) estimate that the effect of interlopers is to reduce $r_0$ by up to $\sim$ 7\% in the BX and BM intermediate-luminosity samples and essentially by zero at $z \sim 3$ --- a negligible effect in all three cases.  In the $G{\mathcal R}I$-selected $z \sim 4$ sample the interloper fraction can be higher --- up to 20\% (Steidel et al.\ 1999) --- but while this likely affects the value of $r_0$, it is unlikely to affect the uncontroversial luminosity-dependent trend seen at this epoch. 

Finally, the choice of fitting region and bin size also affect the clustering measurements, but these factors are generally found to be negligible compared to the factors mentioned above.

\subsection{UV Magnitude dependence}\label{UVmag.sec} 

\begin{figure}
%\plotone{fig03.ps}
\begin{center}
\includegraphics[width=8cm]{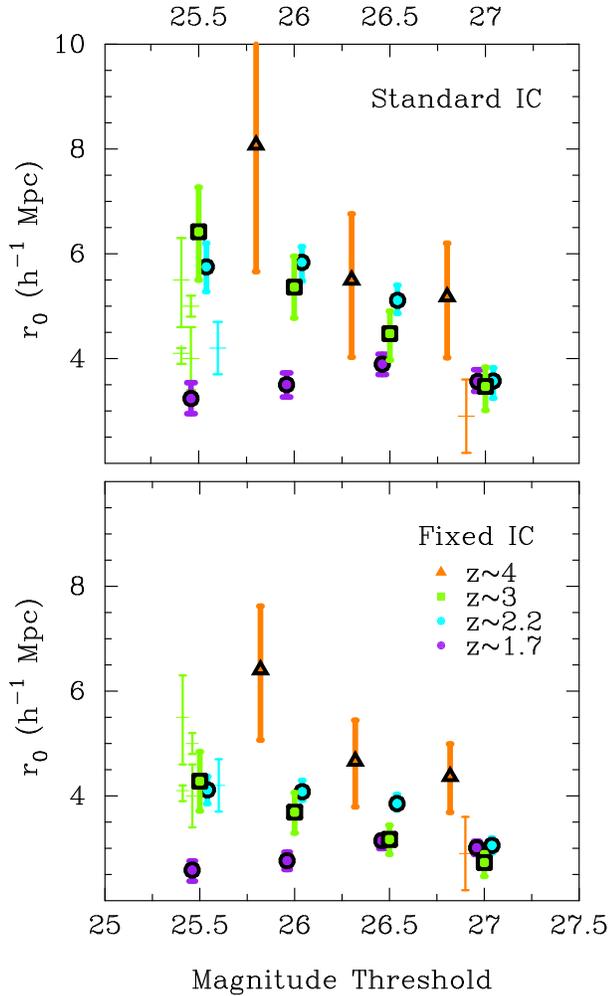}
%\plotone{./figs/r0.ps}
\caption{\label{r0.fig} 
Correlation length, r$_0$, with $\beta$=0.6 for various magnitude thresholds in $\mathcal{R}$ band for the BM, BX and UGR samples and in the $I$ band for the GRI sample.  The top panel is for the standard IC fitting procedure and the bottom panel is for the fixed IC fitting procedure (fit between 20 and 200'' for the BX and UGR samples and 20-150'' for the GRI sample).  Plus signs indicate the results of previous studies discussed in the text.  The overall trends for each redshift sample are independent of the fitting technique, although there is a systematic offset.}
%\vspace{10pt}}
\end{center}
\end{figure}

\begin{figure}
\begin{center}
%\plotone{fig04.ps}
\includegraphics[width=7cm]{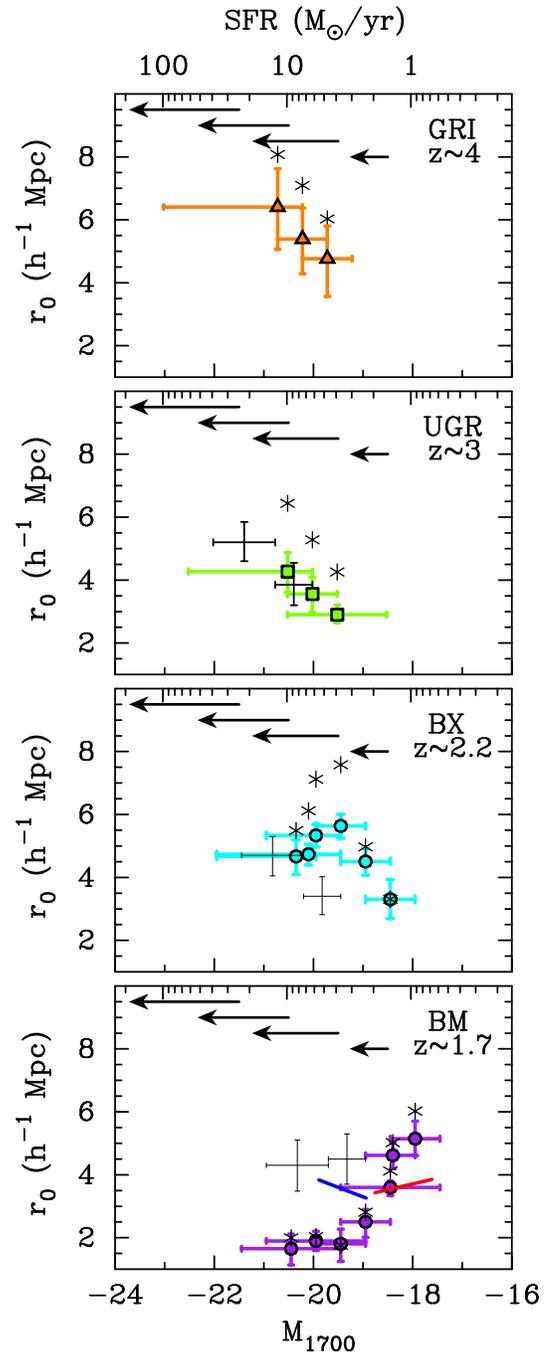}
%\plotone{./figs/r0M1700.ps}
\caption{\label{r0M1700.fig} 
Correlation lengths, r$_0$, for various magnitude ranges using the fixed IC fitting procedure  and $\beta$=0.6.  The top axis corresponds to the implied star formation rate, uncorrected for dust.  Solid arrows represent approximate magnitude-dependent dust corrections for $z\sim2.2$ galaxies from \cite{Sawicki-2011}. The asterisks represent the standard IC fitting results, and the blue and red lines represent the $z\sim1$ and $z\sim0.3$ results of \cite{Heinis-2007}.  The thin black points are those of \cite{Adelberger-2005}.}
%\vspace{10pt}}
\end{center}
\end{figure}

\begin{deluxetable*}{llccccc}
%\tabletypesize{\scriptnotesize}
\tablewidth{0pt} 
\tablecaption{\label{CF.tab}Correlation function parameters for UV-selected samples.}
%\rotate
\tablehead{
\colhead{Sample} &
\colhead{mag} & 
\colhead{$N_{gal}$} & 
\colhead{IC} &
\colhead{A$_w$} &
\colhead{r$_0$} & 
\colhead{r$_{0}$\tablenotemark{a}} \\ %\tablenotemark{a}
\colhead{} &
\colhead{} & 
\colhead{} & 
\colhead{} &
\colhead{[$\beta$ = 0.6]} &
\colhead{[$h^{-1}$ Mpc]} & 
\colhead{[$h^{-1}$ Mpc]} 
}
\startdata
GRI ($z\sim4$) ...…...........…	&	23.0	$\le$	$I$	$\le$	26.8	&	323	& 	0.049	&	1.05	$\pm$	0.35	&	5.2	$\pm$	1.2	&	4.4	$\pm$	0.7	\\
	&	23.0	$\le$	$I$	$\le$	26.3	&	232	& 	0.058	&	1.25	$\pm$	0.49	&	5.5	$\pm$	1.5	&	4.7	$\pm$	0.9	\\
	&	23.0	$\le$	$I$	$\le$	25.8	&	120	& 	0.118	&	2.59	$\pm$	1.08	&	8.1	$\pm$	2.4	&	6.4	$\pm$	1.3	\\
	&						&		& 		&				&				&				\\
UGR ($z\sim3$) .............…	&	23.0	$\le$	$\mathcal{R}$	$\le$	27.0	&	1402	& 	0.018	&	0.40	$\pm$	0.08	&	3.5	$\pm$	0.5	&	2.7	$\pm$	0.3	\\
	&	23.0	$\le$	$\mathcal{R}$	$\le$	26.5	&	1050	& 	0.028	&	0.63	$\pm$	0.11	&	4.5	$\pm$	0.5	&	3.2	$\pm$	0.3	\\
	&	23.0	$\le$	$\mathcal{R}$	$\le$	26.0	&	709	& 	0.039	&	0.73	$\pm$	0.16	&	5.4	$\pm$	0.6	&	3.7	$\pm$	0.4	\\
	&	23.0	$\le$	$\mathcal{R}$	$\le$	25.5	&	424	& 	0.055	&	1.23	$\pm$	0.27	&	6.4	$\pm$	0.9	&	4.3	$\pm$	0.7	\\
	&						&		& 		&				&				&				\\
BX ($z\sim2.2$) ............….	&	23.0	$\le$	$\mathcal{R}$	$\le$	27.0	&	2312	& 	0.015	&	0.35	$\pm$	0.05	&	3.6	$\pm$	0.3	&	3.1	$\pm$	0.1	\\
	&	23.0	$\le$	$\mathcal{R}$	$\le$	26.5	&	1910	& 	0.028	&	0.65	$\pm$	0.06	&	5.1	$\pm$	0.3	&	3.9	$\pm$	0.2	\\
	&	23.0	$\le$	$\mathcal{R}$	$\le$	26.0	&	1483	& 	0.037	&	0.84	$\pm$	0.08	&	5.8	$\pm$	0.4	&	4.1	$\pm$	0.2	\\
	&	23.0	$\le$	$\mathcal{R}$	$\le$	25.5	&	1020	& 	0.038	&	0.86	$\pm$	0.11	&	5.8	$\pm$	0.5	&	4.1	$\pm$	0.3	\\
	&						&		& 		&				&				&				\\
BM ($z\sim1.7$) ............…	&	23.0	$\le$	$\mathcal{R}$	$\le$	27.0	&	1966	& 	0.026	&	0.59	$\pm$	0.05	&	3.6	$\pm$	0.2	&	3.0	$\pm$	0.1	\\
	&	23.0	$\le$	$\mathcal{R}$	$\le$	26.5	&	1707	& 	0.033	&	0.76	$\pm$	0.06	&	3.9	$\pm$	0.2	&	3.1	$\pm$	0.1	\\
	&	23.0	$\le$	$\mathcal{R}$	$\le$	26.0	&	1330	& 	0.034	&	0.76	$\pm$	0.08	&	3.5	$\pm$	0.2	&	2.8	$\pm$	0.2	\\
	&	23.0	$\le$	$\mathcal{R}$	$\le$	25.5	&	971	& 	0.035	&	0.79	$\pm$	0.11	&	3.2	$\pm$	0.3	&	2.6	$\pm$	0.2	\\
\hline																							
\hline																							
	&						&		& 		&				&				&				\\
GRI ($z\sim4$) ...…...........…	&	23.0	$\le$	$I$	$\le$	25.8	&	120	& 	0.118	&	2.49	$\pm$	1.08	&	8.1	$\pm$	2.4	&	6.4	$\pm$	1.3	\\
	&	25.3	$\le$	$I$	$\le$	26.3	&	179	& 	0.082	&	1.77	$\pm$	0.64	&	7.1	$\pm$	1.7	&	5.4	$\pm$	1.1	\\
	&	25.8	$\le$	$I$	$\le$	26.8	&	201	& 	0.056	&	1.18	$\pm$	0.56	&	6.0	$\pm$	2.0	&	4.8	$\pm$	1.2	\\
	&						&		& 		&				&				&				\\
UGR ($z\sim3$) .............…	&	23.0	$\le$	$\mathcal{R}$	$\le$	25.5	&	425	& 	0.055	&	1.23	$\pm$	0.09	&	6.4	$\pm$	0.9	&	4.3	$\pm$	0.7	\\
	&	25.0	$\le$	$\mathcal{R}$	$\le$	26.0	&	513	& 	0.037	&	0.82	$\pm$	0.22	&	5.3	$\pm$	0.9	&	3.6	$\pm$	0.6	\\
	&	25.0	$\le$	$\mathcal{R}$	$\le$	27.0	&	1206	& 	0.024	&	0.53	$\pm$	0.27	&	4.3	$\pm$	0.5	&	2.9	$\pm$	0.3	\\
	&						&		& 		&				&				&				\\
BX ($z\sim2.2$) ............….	&	23.0	$\le$	$G\mathcal{R}$	$\le$	25.0	&	561	& 	0.037	&	0.83	$\pm$	0.21	&	5.5	$\pm$	0.9	&	4.7	$\pm$	0.6	\\
	&	23.0	$\le$	$G\mathcal{R}$	$\le$	25.5	&	943	& 	0.042	&	0.95	$\pm$	0.12	&	6.1	$\pm$	0.5	&	4.7	$\pm$	0.3	\\
	&	24.0	$\le$	$G\mathcal{R}$	$\le$	25.5	&	838	& 	0.052	&	1.19	$\pm$	0.14	&	7.1	$\pm$	0.5	&	5.3	$\pm$	0.4	\\
	&	25.0	$\le$	$G\mathcal{R}$	$\le$	26.0	&	833	& 	0.052	&	1.19	$\pm$	0.14	&	7.6	$\pm$	0.6	&	5.6	$\pm$	0.4	\\
	&	25.5	$\le$	$G\mathcal{R}$	$\le$	26.5	&	906	& 	0.025	&	0.55	$\pm$	0.13	&	5.0	$\pm$	0.7	&	4.5	$\pm$	0.4	\\
	&	26.0	$\le$	$G\mathcal{R}$	$\le$	27.0	&	874	& 	0.012	&	0.25	$\pm$	0.13	&	3.3	$\pm$	1.2	&	3.3	$\pm$	0.6	\\
	&						&		& 		&				&				&				\\
BM ($z\sim1.7$) ............…	&	23.0	$\le$	$G$	$\le$	25.0	&	486	& 	0.020	&	0.42	$\pm$	0.25	&	2.0	$\pm$	0.9	&	1.7	$\pm$	0.5	\\
	&	23.5	$\le$	$G$	$\le$	25.5	&	795	& 	0.017	&	0.37	$\pm$	0.14	&	2.0	$\pm$	0.5	&	1.9	$\pm$	0.3	\\
	&	24.5	$\le$	$G$	$\le$	25.5	&	576	& 	0.013	&	0.26	$\pm$	0.16	&	1.8	$\pm$	0.8	&	1.8	$\pm$	0.6	\\
	&	25.0	$\le$	$G$	$\le$	26.0	&	703	& 	0.019	&	0.41	$\pm$	0.16	&	2.8	$\pm$	0.7	&	2.5	$\pm$	0.5	\\
	&	25.0	$\le$	$G$	$\le$	27.0	&	1390	& 	0.025	&	0.56	$\pm$	0.08	&	4.1	$\pm$	0.4	&	3.6	$\pm$	0.3	\\
	&	25.5	$\le$	$G$	$\le$	26.5	&	744	& 	0.032	&	0.72	$\pm$	0.16	&	5.0	$\pm$	0.7	&	4.6	$\pm$	0.4	\\
	&	26.0	$\le$	$G$	$\le$	27.0	&	683	& 	0.032	&	0.72	$\pm$	0.16	&	6.0	$\pm$	0.9	&	5.1	$\pm$	0.5	\\
\enddata
\tablenotetext{a}{Correlation length for fixed IC value of 0.01.}
\end{deluxetable*}

Previous studies \citep[e.g.,][]{Adelberger-2005, Lee-2006, Kashikawa-2006} indicate that the clustering strength of UV-selected high-$z$ galaxies is dependent on UV magnitude, with brighter galaxies clustering more strongly, but many of these studies have not sampled objects significantly fainter than $M^*$ ($\sim-21$).  The right-hand panels of Figure~\ref{wlsall.fig} show the correlation functions for each redshift sample with different magnitude thresholds.  Table~\ref{CF.tab} and Figures~\ref{r0.fig} and~\ref{r0M1700.fig} summarize the clustering parameters for the two IC fitting techniques.  The results are presented for various magnitude thresholds, as well as various magnitude ranges at 1700 \AA\ for each sample that will be used to investigate the correspondence between dark matter halo mass and UV magnitude in Section~\ref{DM-UV.sec}.

At $z \sim$ 3 and 4, the previous trend of increasing clustering strength with increasing UV magnitude \citep[e.g..][]{Adelberger-2005, Lee-2006, Kashikawa-2006} is confirmed and extended to sub-$L^*$ galaxies.  At $z \sim$ 2.2, this trend is only observed for objects fainter than $\mathcal{R}\sim$ 25 (see Table~\ref{CF.tab}, Fig.~\ref{r0M1700.fig}).  It is interesting to note that the most clustered objects at $z \sim$ 2.2 are not the brightest ones, but those with $\mathcal{R}=$ 25-26 (Fig.~\ref{r0M1700.fig}).  This is in disagreement with the findings of \cite{Adelberger-2005} for objects with $\mathcal{R}<$ 25, but it is in agreement with the findings of \cite{Quadri-2007}, who report that for their sample of $K$-selected galaxies at $z\sim2.5$, those with $R>$ 25 cluster more strongly than those with $R<$ 25.  Our inclusion of fainter optically selected objects reveals an unexpected similarity in the clustering properties of $K$-selected galaxies and optically selected galaxies at $z \sim$ 2.2.  At $z\sim 1.7$, the UV magnitude trend appears to be reversed with respect to $z\sim3$ and 4, with the faintest galaxies being the most strongly clustered.  It should be noted that while the numerical values of $r_0$ vary, the overall trends in each sample are independent of the IC fitting technique.  We discuss the potential implications of these trends with redshift in the context of galaxy downsizing in Section~\ref{downsizing.sec}.

\subsection{UV color dependence}\label{colrel.sec}

\begin{figure}
%\plotone{./figs/colrel.ps}
%\plotone{fig05.ps}
\begin{center}
\includegraphics[width=8cm]{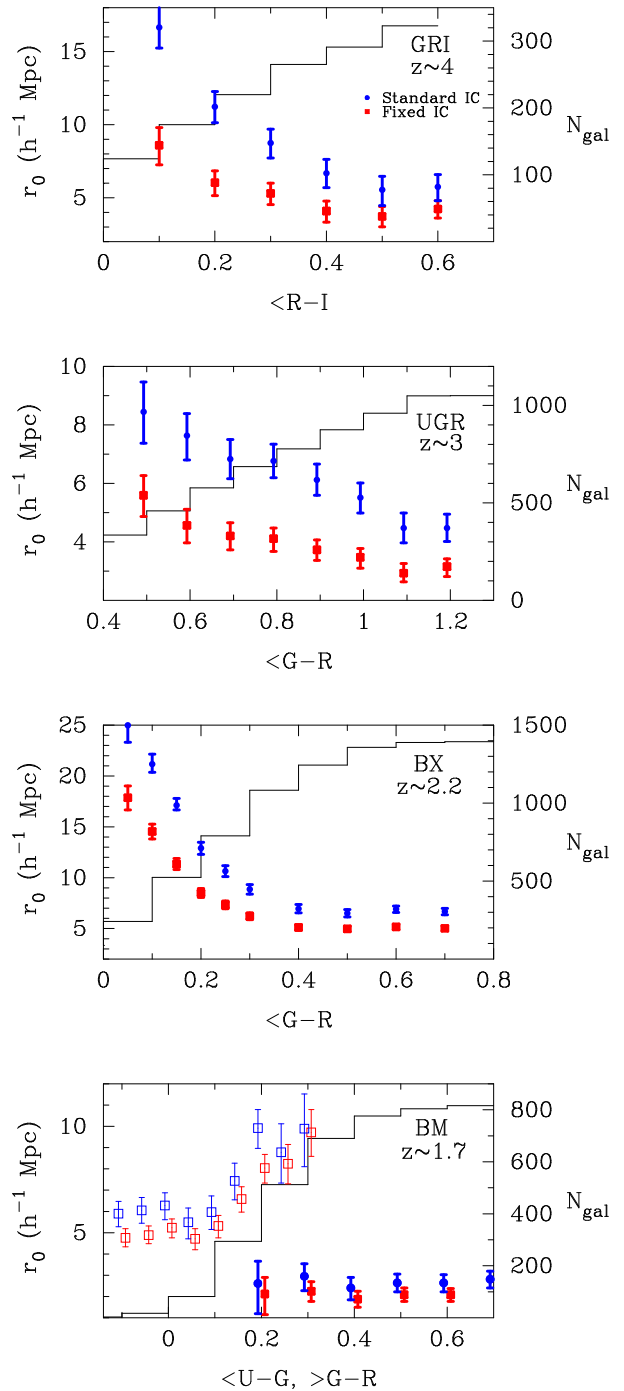}
\caption{\label{colrel.fig} 
Correlation lengths (with $\beta$=0.6) for all galaxies with M$_{1700}\le-19$ and UV colors that are bluer than a given threshold.  Blue circles and red squares correspond to the standard and fixed IC fitting procedures.  Histograms indicate the cumulative number of galaxies below the color threshold and M$_{1700}\le-19$.  The open points in the $z\sim1.7$ panel are for all galaxies with $-20\le M_{1700}\le-19$ redder than a given threshold in $G-\mathcal{R}$. }
%\vspace{10pt}}
\end{center}
\end{figure}

One relationship that has not been previously reported is an association between the correlation length of a sample and its UV color.  Figure~\ref{colrel.fig} shows the relation between sample UV color threshold and correlation length for galaxies with M$_{1700}\le-19$.  At $z\sim$ 4, 3, and 2.2, the galaxies with blue UV colors are significantly more clustered.  There does not seem to be a relation among UV-bright galaxies at $z\sim 1.7$, but faint galaxies with UV red colors show evidence of being more strongly clustered.  This apparent UV color-dependent clustering is independent of the IC fitting technique, and provides a caution against making direct comparisons between studies that use different selection techniques, or between galaxy samples at different redshifts, where the exact \emph{color} selections are not identical.  We note that finding redder galaxies to be weakly clustered seems counterintuitive, and we discuss the possibility that this trend is due to differences in dust properties in Section~\ref{dust.sec} 

The especially striking UV color relationship at $z\sim$ 2.2, in which the measured correlation lengths for both IC fitting techniques increase to well in excess of 10 $h^{-1}$ Mpc (for the bluest samples) may seem unusual for UV-selected galaxies, but we are reminded of the similarities mentioned in the previous section to $K$-selected galaxies, which are routinely observed to have correlation lengths of these magnitudes \citep{Quadri-2007}. Taken together with the observed magnitude dependence of the clustering, this provides evidence of significant overlap between $z\sim$ 2.2 optically selected samples and those obtained using $K$-selection, which are generally thought to be very different populations due to differences in clustering strength.

\subsection{Comparison with Other Surveys}
There are still discrepancies between various surveys concerning the clustering properties of high redshift galaxies, which are generally attributed to differences in selection techniques.  This suggestion is supported by the evidence presented in Section~\ref{colrel.sec} that indicates that in addition to the UV magnitude of objects, UV color can also significantly affect the observed clustering properties. 

One advantage of the KDF survey is that it shares the same selection criteria as work by the Steidel et al. group, a fact that allows us to make comparisons that are likely independent of selection differences.  For the same  selection criteria as those used here, \cite{Adelberger-2005} report correlation lengths of 4.2 $\pm$ 0.5 and 4.0 $\pm$ 0.6 $h^{-1}$ Mpc for galaxies at $z \sim$ 2.2 and 3 with $\gamma \sim$ 1.6 and $\mathcal{R}$ = 23.5-25.5.   Despite the fact that we have identical selection techniques, when the KDF data of similar magnitudes are fit with the standard IC fitting procedure, we obtain larger results 5.8 $\pm$ 0.5 and 6.4 $\pm$ 0.9 $h^{-1}$ Mpc.  The discrepancy can be reduced if force smaller IC values in our analysis. If the fixed IC fitting technique is used, the lower IC value reduces the correlation lengths to 4.1 $\pm$ 0.3 and 4.3 $\pm$ 0.7 $h^{-1}$ Mpc, which are consistent with those of \cite{Adelberger-2005}.  This emphasizes the importance of the IC.  Also worth mentioning is that for galaxies with the same magnitude range at $z \sim$ 3, \cite{Hildebrandt-2007} and \cite{Yoshida-2008} report values of 5.0 $\pm$ 0.2 and 5.5 $\pm$ 0.9 $h^{-1}$ Mpc , which are larger than those of \cite{Adelberger-2005}, but still slightly less than our standard IC fitting results.

For the GOODS fields, \cite{Lee-2006} report clustering measurements for galaxies at $z \sim$ 3 and 4. At $z \sim$ 3 they report a value of $r_0$=4.0 $\pm$ 0.2 $h^{-1}$ Mpc, similar to the value reported in \cite{Adelberger-2005}.  Our larger results for the standard IC fitting procedure can again be traced to large IC values.  At $z \sim$ 4, \cite{Lee-2006} obtained a deep sample with $z_{850} \le$ 27, and report a correlation length of 2.9 $\pm$ 0.7 $h^{-1}$ Mpc.  For a similar magnitude cut ($I  \le$ 26.8), we obtain a value of 5.2 $\pm$ 1.2 $h^{-1}$ Mpc with the standard IC fitting procedure and a value of 4.4 $\pm$ 1.2 $h^{-1}$ Mpc for the fixed IC fitting procedure.  This difference could possibly be caused by a difference in sample selection.  In addition to being magnitude selected in different filters, it is also possible that the Lee et al. sample contains redder objects which are shown to cluster much less strongly in Figure~\ref{colrel.fig}.  In the KDFs, objects in the $z \sim$ 4 sample with $\mathcal{R}-I \ge$ 0.1 are essentially not clustered, and they significantly reduce the correlation length for the total sample.  Another difference between the samples is that at the limit of $z_{850} \le$ 27, the GOODS data are estimated to be $\sim$10\% complete, while the KDFs are estimated to be $\sim$50\% complete at $I  \le$ 26.8.

The KDF results at $z \sim$ 3 and 4 are in good agreement with those of the Subaru Deep Field (SDF).  \cite{Kashikawa-2006} obtained a $z \sim$ 4 sample with a limiting magnitude of $i$=27.43.  They report single power law fits to measurements on all scales, including $r \le 20''$.  If our fitting method is altered to include all scales, and we assume the same values of $\beta$ that they report, the KDF results for the standard fitting technique agree well with their findings.  As previously mentioned, the $z \sim$ 3 results from the SDF, with a limiting magnitude of $z'$ = 25.5 \citep{Yoshida-2008}, also agree with our results.

The correlation lengths reported by \cite{Ouchi-2005} for $z \sim$ 4 objects, and those of  \cite{Hildebrandt-2009} at $z \sim$ 3 and 4 are generally consistent with our results for the fixed IC fitting procedure, but are smaller than our results for the standard IC fitting procedure.

In summary, the KDF clustering measurements are generally found to be in good agreement with previous studies, allowing for differences in selction and fitting techniques.

\section{COMPARISON WITH MILLENNIUM SIMULATION}

The spatial correlation function of dark matter halos is known to be a function of halo mass \citep[e.g.,][]{Mo-1996}, and it has also been established that the correlation functions of UV-selected galaxies are dependent on UV magnitude \citep[e.g.,][]{Adelberger-2005, Lee-2006, Kashikawa-2006}.  In order to investigate a possible relationship between UV magnitude and halo mass, we use the publicly-available Millennium Simulation halo catalogs to estimate the spatial correlation function of halos of various mass, and relate them to our observations of UV-selected galaxies.  

\begin{figure}
%\plotone{./figs/mil_thresh.ps}
\begin{center}
\includegraphics[width=7.5cm]{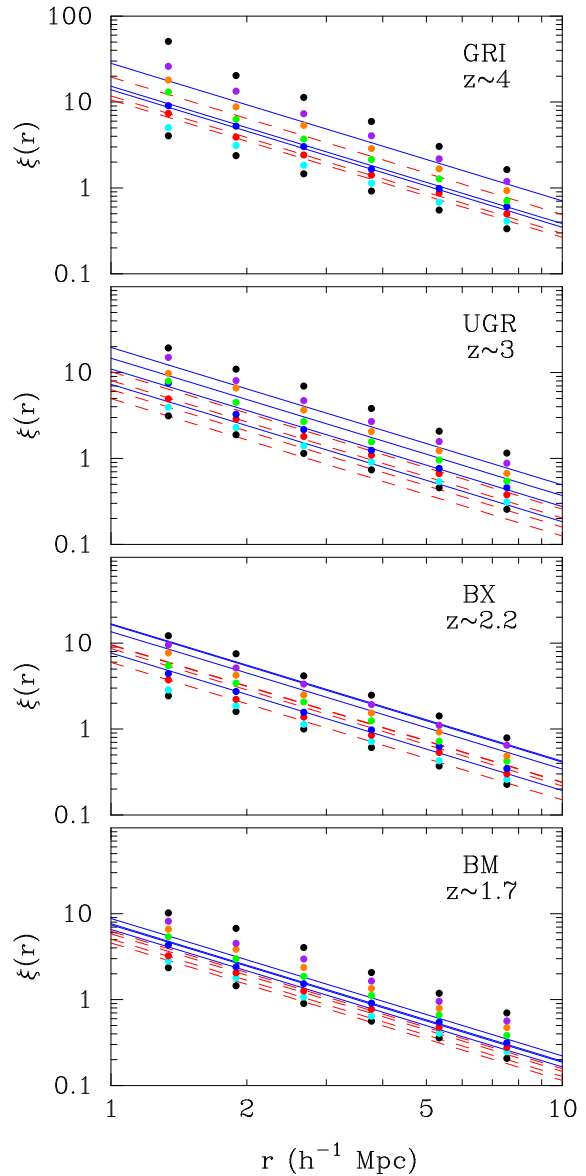}
%\plotone{fig06.ps}
\caption{\label{mil_thresh.fig} 
Comparison of observed spatial correlation lengths for various UV magnitude thresholds (summarized in Table~\ref{CF.tab}) to the predictions of the Millennium Simulation for different halo mass thresholds.  Filled points correspond to increasing halo mass thresholds of log($h^{-1}$M$_\odot$)=10.7, 10.9, 11.1, 11.3, 11.5, 11.7, 11.9, 12.1.  Solid lines represent the results of the standard IC fitting procedure, and the dashed lines are for the fixed IC fitting procedure.
\vspace{10pt}}
\end{center}
\end{figure}

\begin{figure}
%\plotone{./figs/mil.ps}
%\plotone{fig07.ps}
\begin{center}
\includegraphics[width=7.5cm]{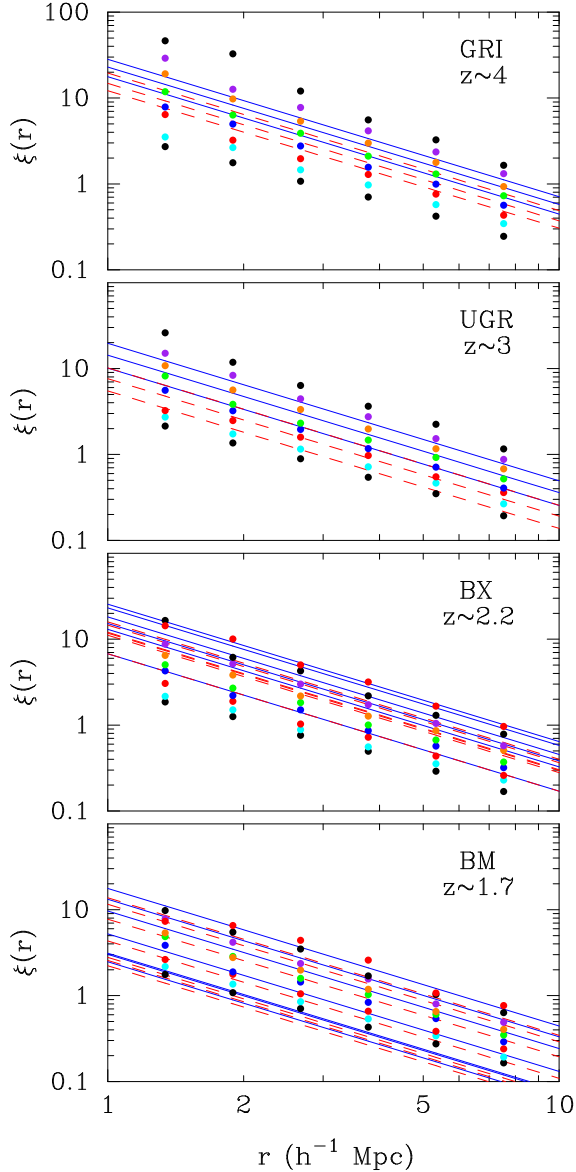}
\caption{\label{mil.fig} 
Comparison of spatial correlation lengths of UV luminosity ranges (summarized in Table~\ref{CF.tab}) to the predictions of the Millennium Simulation for distinct halo mass ranges.  Filled points correspond to increasing halo mass ranges of log($h^{-1}$M$_\odot$)=10.7-11.0, 11.0-11.3, 11.3-11.5, 11.5-11.7, 11.7-11.9, 11.9-12.1, 12.1-12.3, 12.3-12.5.  An additional mass range of log($h^{-1}$M$_\odot$)=12.5-12.7 is shown for the BX and BM samples. Solid lines represent the results of the standard IC fitting procedure, and the dashed lines are for the fixed IC fitting procedure.}
\end{center}
\end{figure}

The dependence of halo clustering on mass in the Millennium simulation is investigated in two manners.  In one method, all haloes with masses greater than a given \emph{threshold} are selected, and for the second method, all haloes within a given mass \emph{range} are selected.  The first method enables a comparison between the number densities of haloes and observed galaxies, and the second is used to develop a relationship between UV magnitude and a typical halo mass.  The results of the two techniques are illustrated in Figure~\ref{mil_thresh.fig} and Figure~\ref{mil.fig}, respectively, and compared to the observed clustering of galaxies with various UV magnitude thresholds and ranges.  To exclude the effects of multiple galaxies residing in the same halo, the comparison is made on large scales ($1\le r \le 10$ $h^{-1}$ Mpc). Given the large uncertainties, rough halo mass estimates are established by eye for both techniques, rather than implementing a formal fitting procedure.  In cases where the assumed slope of $\beta$=0.6 for the observed galaxies does not agree with the predictions of the simulation, a slope of 0.4 or 0.8 is also investigated.  The quoted uncertainties for all mass estimates are determined by plotting the associated correlation function uncertainties over the simulation results in the same manner as in Figures~\ref{mil_thresh.fig} and \ref{mil.fig}.

\subsection{Halo Occupation Numbers}

One advantage of comparing our clustering measurements to the distribution of dark matter haloes in the Millennium Simulation is that it allows us to make comparisons between the observed number densities of galaxies and the number densities of haloes with similar correlation functions.  The comparison is characterized by the mean halo occupation number, $N_{occ}$, which is defined as the ratio of the galaxy number density to that of the haloes.  The number densities of galaxies sampled by the KDFs are found by integrating the observed luminosity functions presented in \cite{Sawicki-2006} down to a given UV magnitude threshold.  The correlation functions for these galaxies are then compared to the predictions of the Millennium simulation for various threshold masses (Figure~\ref{mil_thresh.fig}), which have known number densities. Table~\ref{numdens.tab} summarizes the number densities of observed galaxies and dark matter halos with similar correlation functions.  The ranges reported for the quantity $n_{obs}/n_{halo}$ are found from the maximum and minimum values allowed by the uncertainty in $n_{halo}$.

For most samples, the number of estimated haloes is generally consistent with the number of galaxies, implying that there is about one galaxy for every halo.  This is not the case for the $z\sim$ 2.2 and $z\sim$ 1.7 samples (especially for the standard IC fitting results), where there are more galaxies than haloes by a factor of a few.  We are again reminded of possible similarities to the $K$-selected $z\sim$ 2.2 sample of \cite{Quadri-2007, Quadri-2008}, who report a very large halo occupation number of $\sim$40 for their reddest sample (DRGs, with $J-K>2.3$), and a value of $\sim$1-4 for their complete sample ($K<21)$.

\begin{deluxetable*}{llccccccc}
%\tabletypesize{\scriptnotesize}
\tablewidth{0pt} 
\tablecaption{\label{numdens.tab}Galaxy and halo number densities.}
%\rotate
\tablehead{
\colhead{} &
\colhead{} & 
\colhead{$n_{obs}$} & 
\colhead{Mass} &
\colhead{$n_{halo}$\tablenotemark{a}} &
\colhead{}  &
\colhead{Mass} &
\colhead{$n_{halo}$\tablenotemark{b}} &
\colhead{}  \\
\colhead{Sample} &
\colhead{mag} & 
\colhead{$(\times10^{-3})$} & 
\colhead{Threshold\tablenotemark{a}} &
\colhead{$(\times10^{-3})$} & 
\colhead{$n_{obs}/n_{halo}$\tablenotemark{a}}    &
\colhead{Threshold\tablenotemark{b}} &
\colhead{$(\times10^{-3})$} & 
\colhead{$n_{obs}/n_{halo}$\tablenotemark{b}}  \\
\colhead{} &
\colhead{} & 
\colhead{[Mpc$^{-3}$]} & 
\colhead{[log($h^{-1}$M$_\odot$)]} &
\colhead{[Mpc$^{-3}$]} &
\colhead{}  &
\colhead{[log($h^{-1}$M$_\odot$)]} &
\colhead{[Mpc$^{-3}$]} &
\colhead{}
}
\startdata %& & & & & & & & \\

GRI...........…	&	23	$\le$	$I$	$\le$	26.8	&	1.5	& 	11.30$^{+0.20}_{-0.30}$	&	1.5$^{+2.2}_{-0.7}$	&	0.4	-	2.0	& 	11.0$^{+0.25}_{-0.30}$	&	3.7$^{+3.9}_{-2.0}$	&	0.2	-	0.9	\\
	&	23	$\le$	$I$	$\le$	26.3	&	1.0	& 	11.35$^{+0.25}_{-0.35}$	&	1.3$^{+2.5}_{-0.7}$	&	0.3	-	1.9	& 	11.1$^{+0.20}_{-0.30}$	&	2.8$^{+4.9}_{-1.3}$	&	0.1	-	0.7	\\
	&	23	$\le$	$I$	$\le$	25.8	&	0.6	& 	11.80$^{+0.20}_{-0.40}$	&	0.3$^{+0.8}_{-0.1}$	&	0.6	-	5.6	& 	11.5$\pm$0.20	&	0.8$^{+0.7}_{-0.4}$	&	0.4	-	1.7	\\
	&						&		& 		&		&				& 		&		&				\\
UGR.........…	&	23	$\le$	$\mathcal{R}$	$\le$	27.0	&	4.8	& 	11.15$\pm$0.15	&	3.8$^{+1.9}_{-2.3}$	&	0.8	-	3.2	& 	$<$10.7	&	$>$11.0	&		$<$	0.4	\\
	&	23	$\le$	$\mathcal{R}$	$\le$	26.5	&	3.2	& 	11.30$\pm$0.10	&	2.5$^{+0.8}_{-1.3}$	&	1.0	-	2.6	& 	$<$10.9	&	$>$6.6	&		$<$	0.5	\\
	&	23	$\le$	$\mathcal{R}$	$\le$	26.0	&	2.0	& 	11.55$\pm$0.15	&	1.2$^{+0.7}_{-0.5}$	&	1.1	-	2.6	& 	11.0$^{+0.10}_{-0.20}$	&	5.7$^{+2.9}_{-1.3}$	&	0.2	-	0.5	\\
	&	23	$\le$	$\mathcal{R}$	$\le$	25.5	&	1.2	& 	11.80$^{+0.10}_{-0.25}$	&	0.6$^{+0.7}_{-0.2}$	&	1.0	-	3.0	& 	11.3$^{+0.15}_{-0.25}$	&	2.5$^{+2.5}_{-0.9}$	&	0.2	-	0.7	\\
	&						&		&		&		&				&		&		&				\\
BX............…	&	23	$\le$	$\mathcal{R}$	$\le$	27.0	&	7.5	& 	11.35$^{+0.15}_{-0.05}$	&	3.3$^{+0.5}_{-1.1}$	&	2.0	-	3.3	& 	10.90$\pm$0.10	&	9.0$^{+2.3}_{-1.2}$	&	0.7	-	1.0	\\
	&	23	$\le$	$\mathcal{R}$	$\le$	26.5	&	5.6	& 	11.75$^{+0.15}_{-0.05}$	&	1.2$^{+0.2}_{-0.4}$	&	4.2	-	7.3	& 	11.4$^{+0.05}_{-0.10}$	&	2.9$^{+0.8}_{-0.4}$	&	1.5	-	2.2	\\
	&	23	$\le$	$\mathcal{R}$	$\le$	26.0	&	3.9	& 	11.90$\pm$0.10	&	0.8$\pm$0.2	&	3.9	-	6.9	& 	11.45$^{+0.05}_{-0.15}$	&	2.6$^{+1.2}_{-0.3}$	&	1.0	-	1.7	\\
	&	23	$\le$	$\mathcal{R}$	$\le$	25.5	&	2.6	&	11.90$\pm$0.10	&	0.8$\pm$0.2	&	2.5	-	4.5	&	11.45$^{+0.05}_{-0.15}$	&	2.6$^{+1.2}_{-0.3}$	&	0.7	-	1.1	\\
	&						&		& 		&		&				& 		&		&				\\
BM............…	&	23	$\le$	$\mathcal{R}$	$\le$	27.0	&	24.1	& 	11.30$\pm$0.10	& 	4.0$^{+1.1}_{-0.9}$	& 	4.7	-	7.6	& 	10.90$\pm$0.10	& 	10.2$^{+2.6}_{-2.1}$	& 	1.9	-	3.0	\\
	&	23	$\le$	$\mathcal{R}$	$\le$	26.5	&	19.4	& 	11.50$\pm$0.10	& 	2.5$^{+0.7}_{-0.6}$	& 	6.1	-	10.1	& 	11.10$\pm$0.10	& 	6.4$^{+1.7}_{-1.4}$	& 	2.4	-	3.8	\\
	&	23	$\le$	$\mathcal{R}$	$\le$	26.0	&	14.7	& 	11.30$\pm$0.10	& 	4.0$^{+1.1}_{-0.9}$	& 	2.9	-	4.6	& 	10.80$\pm$0.10	& 	12.8$^{+3.1}_{-2.6}$	& 	0.9	-	1.4	\\
	&	23	$\le$	$\mathcal{R}$	$\le$	25.5	&	10.2	& 	11.10$\pm$0.10	& 	6.4$^{+1.7}_{-1.4}$	& 	1.3	-	2.0	& 	10.70$\pm$0.10	& 	15.8$^{+3.7}_{-3.1}$	& 	0.5	-	0.8	\\
																						
\enddata
\tablenotetext{a}{Standard IC fitting procedure.}
\tablenotetext{b}{Fixed IC fitting procedure.}
\end{deluxetable*}

%\vspace{10mm}

\subsection{Relation Between Halo Mass and SFR}\label{DM-UV.sec}

\begin{figure*}
%\plotone{./figs/DMreltot.ps}
\plotone{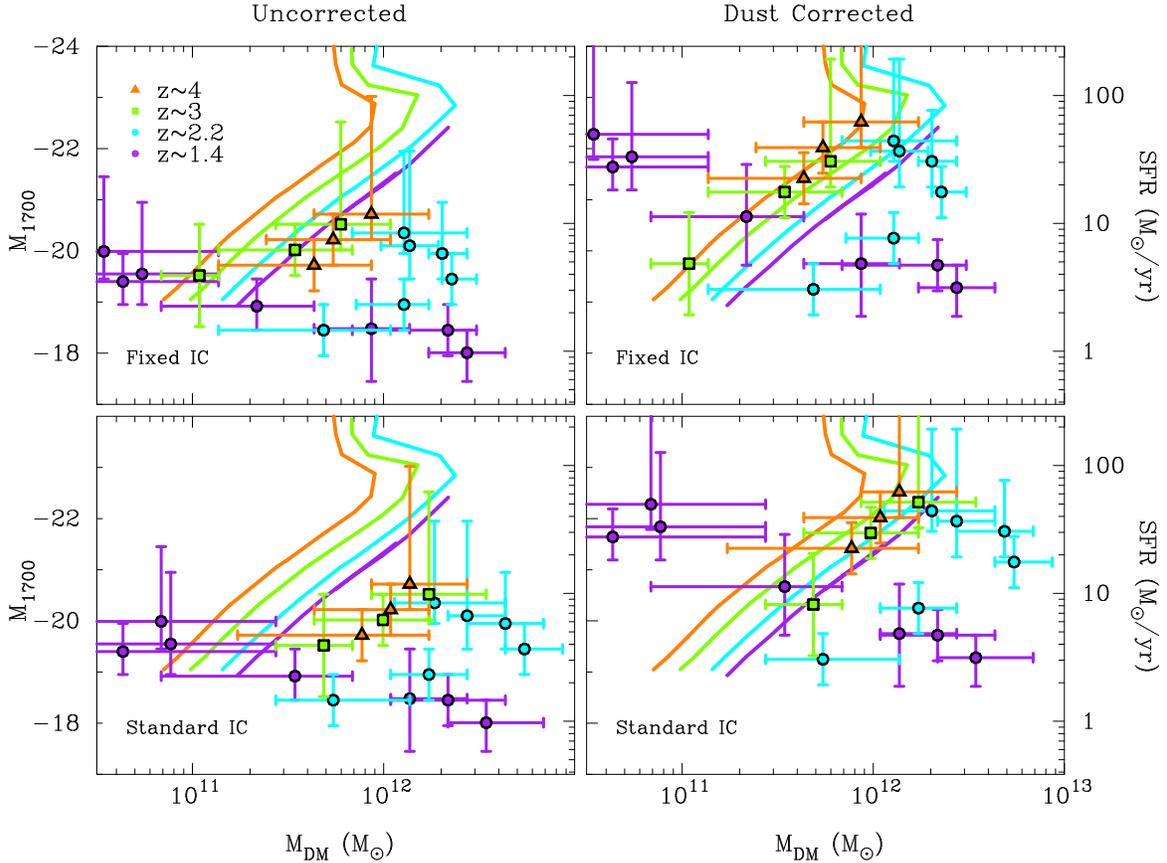}
\caption{\label{DMreltot.fig} 
Relation between dark matter halo mass and UV luminosity/SFR (left and right axes) for all redshift samples.  The bottom panels correspond to the standard IC fitting procedure and the top panels to the fixed IC fitting procedure.  The panels on the left are the observed relations, and the panels on the right are the results after corrections for dust attenuation have been made.  The solid lines correspond to the predictions from the Millennium Simulation galaxy models \citep{DeLucia-2007}, reduced to median relations.}
%\vspace{20pt}}
\end{figure*}

Comparing the observed correlation functions of objects with distinct UV magnitude ranges to the predicted clustering properties of distinct halo masses (Figure~\ref{mil.fig}) allows us to investigate UV luminosities and SFRs associated with halos of a given mass, as a function of redshift.  The left-hand panels of Figure~\ref{DMreltot.fig} show, for the first time, the estimated relationships for galaxies at $z \sim$ 1.7, 2.2, 3, and 4 for the standard and fixed IC fitting procedures (neglecting dust absorption).   It can be seen that for both IC fitting techniques, at $z\sim$3 and 4 the dark matter mass-UV magnitude relationships are essentially indistinguishable, but at $z\sim$ 2.2 the relation is significantly below the higher redshift results. This could possibly indicate that star formation was more active at higher redshift in halos of a given mass, which is consistent with the findings of \cite{Lee-2006}, who find that star formation was more efficient at  $z\sim$ 5 than at $z\sim$ 3-4.  It is worth noting that these conclusions are subject to the amount of dust absorption in the galaxies, but they are independent of the IC fitting procedure used.

The predicted relationship between halo mass and UV luminosity from the Millennium Simulation galaxy catalogs \citep{DeLucia-2007} was also investigated.  To make the comparison, the predicted star formation rates of the modeled galaxies were first converted to M$_{1700}$ magnitudes using the following relation \citep{Sawicki-2006}:
 \begin{equation}
\mathrm{SFR}=6.1\times10^{-(8+0.4M_{1700})} M_{\odot} yr^{-1},
\end{equation}
which neglects dust absorption and is appropriate for a stellar population with a \cite{Salpeter-1955} initial mass function over a range of 0.1 $\le M/M_{\odot} \le$ 100.

In order to make comparisons between the simulated galaxies and our results, at each redshift the predictions are reduced to a median relation between halo mass and UV magnitude (represented as solid lines in Figure~\ref{DMreltot.fig}).  While the median values are convenient for making comparisons, it should be noted that the range of halo masses are generally symmetrical in log$(M_{DM})$ and typically have a standard deviation of a factor of $\sim$2 in mass. 

The absence of dust in the conversion is clearly inaccurate, and as expected, the predicted M$_{1700}$ magnitude for galaxies of a given halo mass is much greater than our observed relations.  However, magnitude-dependent dust corrections are now available at $z\sim$ 2.2 from $U$-through-$H$ SED fitting of UV-selected galaxies in the Hubble Deep Field \citep{Sawicki-2007, Sawicki-2011}, and we apply these to our presents KDF results.  It is assumed that the magnitude-dependent corrections at the other redshifts are identical to those at $z\sim$ 2.2.

Accounting for the absorption by dust significantly alters the observed relations (right-hand panels of Figure~\ref{DMreltot.fig}) and brings our results into much better agreement with the predictions.  After the corrections are made, the $z\sim$ 3 and 4 results for both IC fitting procedures agree within the errors with the predictions of the Millennium Simulation.  The agreement is especially striking, given the considerable uncertainties in both the models and our work.  At lower redshifts there are large differences between our results and the model predictions, perhaps reflecting changes in the evolution of galaxies at these redshifts that have not been considered in the model.

\subsection{Stellar to Dark Matter Mass Relation at $z\sim 2.2$}

Another interesting quantity that can be investigated is the relation between stellar and dark matter halo mass, which can be estimated by combining the above relation between halo mass and UV luminosity with a recently developed relation between stellar mass and UV luminosity for star forming galaxies at  $z\sim 2.2$ (\citealt{Sawicki-2011}; see also \citealt{Sawicki-2007}):
\begin{equation}
\label{DMstars.eq}
\log \frac{M_{stars}}{M_{\odot}} = -0.11 - 0.51 M_{1700}, 
\end{equation}
where $M_{1700}$ is the absolute 1700\AA\ luminosity not corrected for dust.  This relation is derived from broadband spectral energy distribution (SED) fitting for a sample of UV-selected $z\sim 2.2$ galaxies extending to $R$=28 in the Hubble Deep Field.  The observed $U_{300}B_{450}V_{606}I_{814}J_{110}H_{160}$ SEDs are compared to the spectral synthesis models of \cite{Bruzual-2003} with a constant star formation rate and a metallicity of 0.2 $Z_\odot$.  The synthetic spectra are reddened using the \cite{Calzetti-2000} starburst extinction curve, and shifted in wavelength and intensity according to the adopted cosmological parameters, which are the same as those used in this study. The relation is valid for the \cite{Salpeter-1955} stellar initial mass function (IMF).

\begin{figure}
%\plotone{./figs/DMstarsM.ps}
\plotone{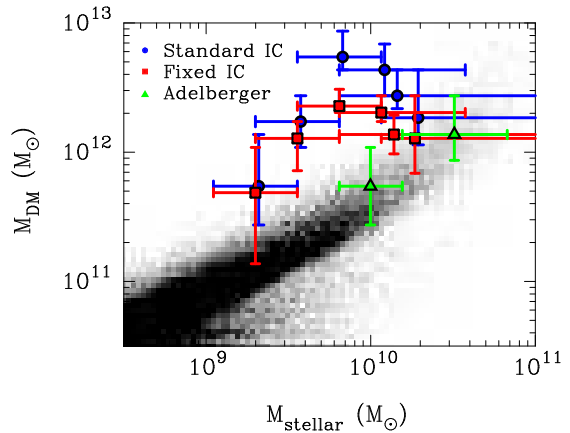}
\caption{\label{DMstars.fig} 
Relation between stellar mass  and dark matter halo mass at $z \sim$ 2.2 for the different IC fitting techniques.  The greyscale represents the predictions from the Millennium Simulation galaxy catalogs.  For comparison, the green triangles show the results of \cite{Adelberger-2005}.
\vspace{10pt}}
\end{figure}

Combining the $z\sim 2.2$ points in Figure~\ref{DMreltot.fig} with Equation~\ref{DMstars.eq} produces the results shown in Figure~\ref{DMstars.fig}.  The two sets of points correspond to the two IC fitting procedures, and the greyscale represents the predictions of the Millennium Simulation galaxy models of \cite{DeLucia-2007}.  There is a significant disagreement, particularly at intermediate halo masses,  between the results for both fitting procedures and the models and our stellar mass estimates are low by a factor of about 5-10, compared to the models. This discrepancy could be indicative of limitations in: (1) our method of estimating dark matter mass from clustering measurements, (2) the SED fitting process used to estimate the stellar masses, or (3) the galaxy models. One possible explanation for the discrepancy is that the total stellar mass is underestimated because it is derived exclusively from observations of the UV continuum which only sample young, recently-formed stars, and do not account for the contribution of existing older stellar populations.   Another possibility is that the estimated dark matter masses are biased towards massive haloes that strongly affect the clustering measurements, while the stellar mass estimates are biased towards the more numerous and less-massive, but still actively star -forming galaxies.   We note, however, that applying Eq.~\ref{DMstars.eq} to the clustering measurements of \cite{Adelberger-2005} gives a reasonably good agreement with the models, albeit this can be tested only at the highest halo masses.

\section{DISCUSSION}

\subsection{Magnitude-Dependent Clustering and Galaxy Downsizing}\label{downsizing.sec} 
The premise of galaxy downsizing originated with the observation that the sites of highest star formation shift from high to low stellar mass systems with time \citep{Cowie-1996}.  A logical extension of this scenario would seem to be that star formation first began in the most massive dark matter haloes and then, as time progressed, the sites of highest star formation migrated from higher to lower mass haloes.  Such an extension of galaxy downsizing from stellar to halo mass is suggested by \cite{Heinis-2007}, who use clustering measurements of UV-bright star forming galaxies at $z\sim$ 1 and 0.2 to determine that these objects reside in significantly lower mass haloes than at $z>$ 2.  Our magnitude-dependent clustering results at $z\sim$ 4, 3, 2.2, and 1.7, are well-suited to extend the framework of galaxy downsizing to halo mass.

At $z\sim$ 3 and 4, the UV-bright galaxies in our sample are observed to cluster more strongly than UV-faint ones, in agreement with previous studies.  We propose that at these redshifts, the current UV-limited samples consist mainly of massive dark matter haloes, in which there is a strong relation between halo mass and SFR (as indicated by UV magnitude).  By $z\sim$ 2.2, a different picture begins to emerge, as we find evidence that the most strongly clustered galaxies in our sample are no longer those with the highest UV luminosities.  By $z\sim$ 1.7, the UV magnitude relation appears to be completely reversed, with the UV-faint galaxies in our sample clustering more strongly than UV-bright ones (Fig.~\ref{r0M1700.fig}). 

Our results suggest that for the objects in our sample, the most massive dark matter haloes host galaxies with high SFRs ($M_{1700}< -20$, or $>50~\rm{M}_\odot$ yr$^{-1}$) at $z\sim$ 3 and 4, moderate SFRs ($-20<M_{1700}< -19$, or $\sim 20~\rm{M}_\odot$ yr$^{-1}$) at $z\sim$ 2.2, and lower SFRs ($-19<M_{1700}< -18$, or $\sim2~\rm{M}_\odot$ yr$^{-1}$) at $z\sim$ 1.7.   The decrease in star formation we observe in the most massive haloes at $z\sim2$ is consistent with the findings of \cite{Hartley-2008} and \cite{Blanc-2008} who report that the population of passively evolving $BzK$-selected galaxies is more strongly clustered than are star-forming galaxies at this redshift. In this scenario, by $z\sim2$, the epoch of large-scale star formation is coming to an end in many of the most massive haloes, which is reflected in the larger correlation lengths of UV-faint galaxies in our sample.   Our results are also consistent with the weak relation between UV magnitude and clustering strength observed by \cite{Heinis-2007} at $z\sim$ 1, because our results suggest that the absolute UV magnitude of the galaxies in the most massive  haloes should be $M_{1700}>-18$, below the threshold of their survey.  The implied shutdown of star formation in the most massive haloes at $1 < z < 2$ is also consistent with the findings of \cite{Arnouts-2007}, who report a buildup of the quiescent red sequence of galaxies between $1 < z < 2$.   The developing consensus that $z\sim2$ is the redshift at which star formation is being quenched in the most massive haloes, while lower mass haloes continue to rapidly form new stars, is further supported by our magnitude-dependent clustering measurements.  These results are also supported by the UV luminosity and SFR densities derived for the KDFs \citep{Sawicki-2006b}, which show that the dominant contributor to the total luminosity changes with redshift, with less luminous galaxies becoming more dominant as time progresses.

It has occurred to us that the number densities of the strongly clustered faint objects in the $z\sim 1.7$ sample are much larger than those of other strongly clustered populations, which is a concern if they are all believed to represent the most massive haloes.  However, we have used results from the Millennium Simulation to determine that the observed $\sim50\%$ increase in clustering for faint objects can be attainted by adding about 1 in 5 high-mass haloes to a sample of weakly clustered low-mass haloes.  It seems plausible that high-mass haloes could account for $\sim 20\%$ of the faint objects in our sample, which could raise the clustering strength by the observed $\sim50\%$.  It should be noted that these estimated increases in clustering strength only apply to very weakly clustered "background" populations and should not be generalized to other populations. 

We believe our clustering measurements provide a new line of evidence for galaxy downsizing.  By using clustering measurements to determine the typical dark matter halo mass of star forming galaxies across a broad range of luminosities and redshifts, we can trace the evolution of the sites of highest star formation (as indicated by UV luminosity) from high-mass haloes at high redshift, to lower-mass haloes at lower redshift.  This suggests an observational link between galaxy downsizing and halo mass. 

The reversal of the UV magnitude trend at $z\sim$ 2 has not been previously noted due to the lack of deep UV coverage needed to select galaxies at these intermediate redshifts.  The very deep coverage of the KDFs  across wavelengths needed to efficiently select galaxies from $z\sim1.5-4.5$, has allowed us to detect this important trend in the evolution of galaxies in the context of halo mass.  The beginnings of this trend were first hinted at by \cite{Quadri-2007}, who report that objects in their $K$-selected sample at $z\sim$ 2.5 with $R>$ 25 cluster more strongly than those with $R<$ 25.  The concept is also supported by \cite{Heinis-2007}, who report that there is very little relation between UV magnitude and clustering strength for galaxies at lower $z$.  Our findings provide a significant confirmation and extension of these results.

\subsection{UV color relation: Observing the 2175 $\rm{\AA}$ UV absorption bump?}\label{dust.sec} 

In Section~\ref{colrel.sec} we report that galaxies with blue UV colors (in $G-\mathcal{R}$ or $\mathcal{R}-I$) appear to cluster more strongly at $z\sim$ 4, 3, 2.2, with the relationship being most striking at $z\sim$ 2.2.  This relationship does not exist at $z\sim$ 1.7 and could even be reversed, with UV-red galaxies possibly clustering more strongly (Figure~\ref{colrel.fig}).  We have verified that this trend also exists in $BzK$-selected galaxies at $z\sim2$ by using the publicly available MUSYC catalogs for the ECDFS field, to confirm that galaxies in that sample with blue UV colors (in $B-R$) also cluster much more strongly (but not for $R-I$).  The trend of objects with blue UV colors clustering more strongly is opposite to what is observed in optical colors.  \cite{Quadri-2007} report that for their $K$-limited sample at $z\sim$ 2.5, optically red galaxies (in $J-K$) cluster much more strongly, and raise the question of whether this effect is due to old stellar populations or dust.  

\begin{figure}
%\plotone{./figs/dust.ps}
\begin{center}
\includegraphics[width=7.5cm]{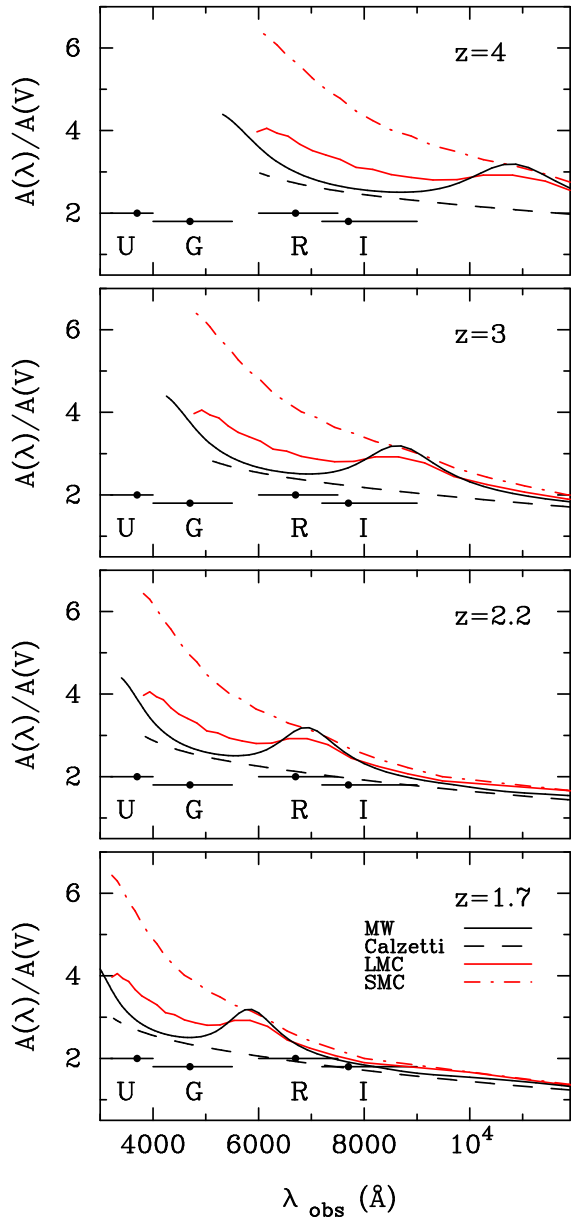}
%\plotone{fig10.ps}
\caption{\label{dust.fig} 
Average extinction curves for local galaxies compared to our $U_nG\mathcal{R}I$ filter set at different sample redshifts. Note the prominence of the 2175 $\rm{\AA}$ absorption bump in Milky Way dust, and its location in the $\mathcal{R}$ band at $z\sim2$.  The data for the LMC and SMC are from \cite{Gordon-2003}, the MW curve (with $R_V=3.1$) is from \cite{Cardelli-1989}, and the starburst curve (with $R_V=4.05$) is from \cite{Calzetti-2000}.}  
%\vspace{10pt}}
\end{center}
\end{figure}

An intriguing possibility is that the strong relation we observe between UV color and clustering strength is a result of the well-known 2175 $\rm{\AA}$ UV absorption bump.  Locally, this broad spectral feature is prominent along most sight-lines in the Milky Way galaxy, is moderately strong in the Large Magellanic Cloud (LMC), and is absent in the Small Magellanic Cloud (SMC) \citep{Gordon-2003} (see Figure~\ref{dust.fig}).  The feature has not been detected in local starburst galaxies \citep{Calzetti-1994}, and in the Milky Way it has been associated with regions of diffuse ISM and seems to be suppressed in dense star-forming areas \citep{Whittet-2004}.  One of the currently favored carriers for the absorption band is polycyclic aromatic hydrocarbons (PAHs), which are produced by AGB stars of intermediate ages.  An age dependence for this absorption feature could explain its absence in areas of recent star formation.   

At $z\sim$ 2.2, this distinct absorption feature has recently been observed in composite spectra of UV-bright galaxies with redder overall UV continuum slopes, but not in those with blue UV slopes \citep{Noll-2005}.  This work has been extended by  \cite{Noll-2007}, who report that at $z\sim$ 2.2, the feature is stronger in more``mature" galaxies, consistent with the carrier of the absorption being produced by older stellar populations.

The argument for this feature playing a role in our UV color relation is especially convincing for the BX sample at $z\sim$ 2.2, where the rest-frame wavelength of 2175 $\rm{\AA}$ falls directly in the $\mathcal{R}$ band (see Figure~\ref{dust.fig}).  If a strong absorption bump is present, it will significantly reduce the $\mathcal{R}$ band flux and actually make the observed $G-\mathcal{R}$ color bluer compared to galaxies without the feature.  Since the absorption bump seems to be associated with relatively evolved stellar populations, it follows that it would be more prominent in more massive haloes at $z\sim$ 2.2, in which star formation is shutting down.  In this scenario, galaxies with relatively blue $G-\mathcal{R}$ colors would then be the most massive and strongly clustered, in agreement with our observations.

In studies of high-$z$ galaxies, it is usually assumed that the dust properties are similar to those found in local starburst galaxies (e.g., \cite{Sawicki-1998, Papovich-2001, Shapley-2001, Shapley-2005, Yabe-2009}).  Since it seems possible that old high-$z$ galaxies contain a Milky Way-type mixture of dust, it will be interesting to consider what effects this could have on global properties, such as SFRs, that are derived from SED-fitting. 

The redshift dependence of the UV-color relations we observe can actually support the concept of galaxy downsizing discussed in the previous section if we believe that the 2175 $\rm{\AA}$ absorption bump is driving the trends and that it is more prominent in massive haloes.  If UV color is an indicator of halo mass (which seems possible, given the strong clustering dependence shown in Figure~\ref{colrel.fig}), our observations can be interpreted in the following way: (1) at  $z\sim$ 3 and 4, there are few low mass haloes in the sample, so the UV color dependence is relatively small, but can still be used to select the most massive haloes.  (2) At  $z\sim$ 2.2, the sites of highest star formation are beginning to migrate from high to low mass haloes, resulting in a significant range of halo masses in the sample, and the ability to separate them based on color produces a very large UV-color relation.  (3) At $z\sim$ 1.7, most galaxies in the sample are hosted by low mass haloes in which the absorption bump is not prominent, and the UV-color relation is removed or reversed, with UV-red galaxies possibly clustering more strongly.  This scenario provides additional support for galaxy downsizing.

Our proposal that the UV-bright objects at $z\sim2$ are hosted in a mixture of high and low mass haloes is consistent with results of \cite{Conroy-2008}, who argue that based on the most up-to-date number density observations, star forming galaxies at $z\sim 2$ (specifically, those with $\mathcal{R}\le25$) cannot evolve into massive red galaxies, as has been previously suggested \citep{Adelberger-2005}.  Taking into account the merging history of $z\sim2$ star forming galaxies, they find that the number densities are most consistent with those of typical $\sim L^*$ galaxies at $z\sim 1$ and 0. Our interpretation allows for a fraction of the $z\sim2$ star forming galaxies (those with a prominent 2175 $\rm{\AA}$ absorption bump and blue $G-R$ colors) to reside in the most massive haloes and possibly evolve into ellipticals by $z\sim0$, while the majority may evolve into typical $L^*$ galaxies like our own.

\section{SUMMARY AND CONCLUSIONS}

We have investigated the clustering properties of UV-selected galaxies at $z\sim$ 4, 3, 2.2, and 1.7, extending to those significantly below $L^*$.  The main findings of our study can be summarized as follows:

\begin{enumerate}

\item  At $z\sim$ 4 and 3, we find that the UV-brightest galaxies are the most strongly clustered, in agreement with previous studies, but this is not the case at $z\sim$ 2.2 and 1.7.  We believe that we are observing galaxy downsizing, as the sites of highest star formation migrate from high mass dark matter haloes at high redshift to lower mass haloes at lower redshift.  

\item By comparing our clustering measurements to the predictions of the Millennium Simulation, we derive halo mass estimates for the galaxies in our samples, and from these comparisons, we develop relationships between halo mass and SFR for each sample.  Our results are consistent with a shut-down of star formation in massive haloes as time progresses.  In particular, we estimate that the the most massive dark matter haloes in our sample ($\sim1-5\times10^{12}~\rm{M}_\odot$) host galaxies with high SFRs ($M_{1700}< -20$, or $>50~\rm{M}_\odot$ yr$^{-1}$) at $z\sim$ 3 and 4, moderate SFRs ($-20<M_{1700}< -19$, or $\sim 20~\rm{M}_\odot$ yr$^{-1}$) at $z\sim$ 2.2, and lower SFRs ($-19<M_{1700}< -18$, or $\sim2~\rm{M}_\odot$ yr$^{-1}$) at $z\sim$ 1.7.  At $z\sim$ 4 and 3, our results are in good agreement with the semi-analytic models of galaxy formation that have been added to the Millennium Simulation, but generally do not agree with the predictions at lower redshifts.

\item The relatively small field of view of the KDFs imposes a systematic dependence on the correlation lengths we infer due to the large integral constraint corrections that need to be applied.  Howerver, by using artificial integral constraint corrections, we have shown that this dependence does not affect any of the important general trends that we observe. 

\item We find that the clustering strength of galaxies in all redshift samples is related to UV color.  We propose that this is due to the presence of the 2175 $\rm{\AA}$ absorption bump in more massive halos, which contain the older stellar populations and dust needed to produce the feature.  This UV color relation provides a caution against comparisons between surveys that have different selection criteria, and also for comparisons between different redshift samples.

\item Our inclusion of fainter UV magnitudes has revealed additional similarities between Lyman break galaxies and $BzK$-selected galaxies at $z\sim 2$, such as $\mathcal{R}$-faint objects clustering more strongly than $\mathcal{R}$-bright ones, and correlation lengths greater than 10 $h^{-1}$ Mpc for subsamples of galaxies selected with each technique. 
%; halo occupation numbers greater than unity; missing stellar mass in our sample

\item Using the Millennium Simulation results, we derive halo occupation numbers for our samples and find that at $z\sim$ 4 and 3, the results are generally consistent with approximately one galaxy per halo.

\item At $z\sim$ 2.2, we estimate the relation between stellar and dark matter mass, and compare it to model predictions.  We find generally poor agreement, and believe this may be due to an underestimate of the stellar masses, whose estimates are dominated by ongoing star formation and likely miss the presence of older stellar populations. This is indeed what one would expect from galaxies in the process of turning down their star formation rates.

\end{enumerate}

\vspace{5mm}

In conclusion, we note that the addition of faint Lyman break galaxies to clustering measurements suggests several interesting new trends; it will be important to extend these findings with new surveys that reach the same faint magnitudes, but cover a larger area, in order to minimize the significance of systematics, in particular those due to the integral constraint.  An additional future direction will be the color-selection of subsamples of the BM, BX, and LBG populations to investigate the effects of the shut-down of star formation in massive haloes.

%acknowledgments
\vspace{8mm}

We wish to recognize and acknowledge the very significant cultural role and reverence that the summit of Mauna Kea has always had within the indigenous Hawaiian community; we are most fortunate to have the opportunity to conduct observations from this mountain. We thank the referee for constructive comments that helped improve this manuscript, and Michael Palmer for producing the greyscale image in Figure~\ref{DMstars.fig}. Parts of the analysis presented here made use of the Perl Data Language (PDL) that has been developed by K.\ Glazebrook, J.\ Brinchmann, J.\ Cerney, C.\ DeForest, D.\ Hunt, T.\ Jenness, T.\ Luka, R.\ Schwebel, and C. Soeller,  which can be obtained from http://pdl.perl.org. PDL provides a high-level numerical functionality for the perl scripting language (Glazebrook \& Economou, 1997). Some of the results presented here are based on observations obtained at the Gemini Observatory, which is operated by the Association of Universities for Research in Astronomy, Inc., under a cooperative agreement with the NSF on behalf of the Gemini partnership: the National Science Foundation (United States), the Science and Technology Facilities Council (United Kingdom), the National Research Council (Canada), CONICYT (Chile), the Australian Research Council (Australia), MinistŽrio da Cincia e Tecnologia (Brazil) and Ministerio de Ciencia, Tecnolog\'ia e Innovaci\'on Productiva  (Argentina). This work was supported by funding from the Natural Sciences and Engineering Research Council of Canada and the Canadian Space Agency.  

%BIBLIOGRAPHY
%REFERENCES
\bibliographystyle{apj}
\bibliography{references}

%\begin{figure}
%\plotone{galcol.ps}
%\plotone{fig09.ps}
%\caption{\label{galcol.fig} 
%Median galaxy color.  
%\vspace{10pt}}
%\end{figure}

%% Generally speaking, only the figure captions, and not the figures
%% themselves, are included in electronic manuscript submissions.
%% Use \figcaption to format your figure captions. They should begin on a
%% new page.

%\clearpage

%% No more than seven \figcaption commands are allowed per page,
%% so if you have more than seven captions, insert a \clearpage

%% There must be a \figcaption command for each legend. Key the text of the
%% legend and the optional \label in curly braces. If you wish, you may
%% include the name of the corresponding figure file in square brackets.
%% The label is for identification purposes only. It will not insert the
%% figures themselves into the document.
%% If you want to include your art in the paper, use \plotone.
%% Refer to the on-line documentation for details.

% FIGURES

%% Two options are available to the author for producing tables:  the
%% deluxetable environment provided by the AASTeX package or the LaTeX
%% table environment.  Use of deluxetable is preferred.
%%

%\clearpage

%\input coad-subsets.tab.tex

%% The following command ends your manuscript. LaTeX will ignore any text

%\newpage
%\clearpage

\end{document}